\documentclass[prb, twocolumn, nofootinbib]{revtex4-2}

\usepackage{bm}
\usepackage{graphicx}
\usepackage[x11names]{xcolor} 
\usepackage{amsmath}
\usepackage{bbold}
\usepackage{mathrsfs}
\usepackage{color}
\usepackage[unicode=true,colorlinks=true]{hyperref}
\usepackage[capitalize]{cleveref}
\newcommand{\KP}{\ensuremath{\bm{k} \cdot \bm{p}}}

\newcommand{\ket}[1]{\left\vert #1 \right\rangle}

\begin{document}

\title{Optical transitions, exciton radiative decay, and valley coherence in lead chalcogenide quantum dots}
\author{S.V.~Goupalov$^{1,2}$} 
\email{serguei.goupalov@jsums.edu}
\author{E.L.~Ivchenko$^{2}$}
\author{M.O.~Nestoklon$^{2}$}
\email{nestoklon@coherent.ioffe.ru}
\affiliation{$^1$~Department of Physics, Jackson State University, Jackson MS 39217, USA\\
$^2$~Ioffe Institute, 194021 St. Petersburg, Russia}

\begin{abstract}
We propose the concept of valley coherence and superradiance in the reciprocal space and show that it leads to an $N$-fold decrease of the bright exciton radiative lifetime in quantum dots (QDs) of an $N$-valley semiconductor. Next we explain why, despite this, the exciton radiative lifetimes in PbX (X = S, Se, Te) QDs, measured from the photoluminescence decay, are in the microsecond range. We also address peculiarities of the light-matter interaction in nanostructures made of narrow-gap materials with strong inter-band coupling.
\end{abstract}

\maketitle

\section{Introduction}
Lead chalcogenide PbX (X = S, Se, Te) quantum dots (QDs) have found numerous applications in optoelectronics~\cite{Sun2012,Sukhovatkin09,Tisdale2010,Gao2020,Lu2020} and {\it in vivo} fluorescence imaging~\cite{Kong2016,Zhang2018,Xia2021} due to tunability of their fundamental optical transition with the QD size within the near-infrared and mid-infrared ranges. Although lead chalcogenides are $N$-valley semiconductors with $N=4$, the simplest models widely used for their description are restricted to the carrier states in a single valley. In the model proposed by Kang and Wise~\cite{Kang1997} electron states in a PbX QD are described by the isotropic two-band Dirac Hamiltonian with a gap term and additional massive terms accounting for the contributions of remote bands. The presence of these massive terms makes relation between the velocity and coordinate matrix elements non-trivial, which may result in a confusion when addressing interband optical transitions. One of the aims of the present work is to clarify this issue. 

It is known from quantum electrodynamics that the operator of the interaction of a charged particle with the electromagnetic field has the universal 
form~\cite{Landau4_book}
\begin{equation}
\hat{V}=\int \rho({\bf r}) \, \varphi({\bf r}) d {\bf r}-\frac{1}{c} \int  {\bf j}({\bf r}) \, {\bf A}({\bf r}) d {\bf r} \,,
\label{V1}
\end{equation}
where $c$ is the speed of light in vacuum, $\rho({\bf r})$ and ${\bf j}({\bf r})$ are, respectively, the charge density and current density operators, $\varphi({\bf r})$ and ${\bf A}({\bf r})$ are the scalar and vector potentials. As far as
the charge and current density operators are not specified, this expression is valid for any charged particle~\cite{Landau4_book}. This statement can be extended to charged quasi-particles
in solids~\cite{Landau9_book}, and Eq.~(\ref{V1}) can be used along with effective Hamiltonians of the \KP~method. Another popular approach in solid state physics is to use the so-called Peierls substitution ${\bf K} \rightarrow {\bf K} -e {\bf A}({\bf r})/\hbar c$ (where $e<0$ is the electron charge and ${\bf K}$ is the quasi-momentum) in the band energy dispersion function $E({\bf K})$ of Bloch electrons or in the multi-band effective Hamiltonian $\hat{H}({\bf K})$ of the effective mass approximation. If this Hamiltonian accounts for the inter-band coupling at non-zero ${\bf K}$ then the Peierls substitution can be used in order to describe optical transitions between these bands. This substitution is justified for weak fields~\cite{Landau9_book,Blount1962}
and, in the linear approximation, leads to the same results as the interaction~(\ref{V1}). If a Bloch electron is described in such a way that the Hamiltonian explicitly contains the periodic potential of the crystal
lattice and the free electron mass $m_0$ is used in the expression for its kinetic energy then, in the gauge
$\varphi({\bf r})=0$, the interaction~(\ref{V1}) is equivalent to $-e \{{\bf A }({\bf r}_e){\bf p}_e\}/c m_0$~\cite{toyozawa},
where ${\bf p}_e$ and ${\bf r}_e$ are the electron momentum and coordinate operators, and curly brackets mean the symmetrized product of operators.
However, when one combines this form of interaction with the Hamiltonian of the effective mass approximation~\cite{Kang1997}, it can result in a confusion.
Finally, one often discusses equivalence of the interaction Hamiltonians $\{ {\bf A} \cdot {\bf p} \}$ and ${\bf E} \cdot {\bf r}$ in the long-wavelength limit (dipole approximation)~\cite{tannoudji1989}. This equivalence holds within the effective mass approximation for coupled bands when the $\{ {\bf A} \cdot {\bf p} \}$ 
is understood in terms of the Peierls substitution.

In the first part of the present work, we consider a fundamental optical transition associated with the ground exciton state in PbX QDs beyond the long-wavelength limit. We use the model of Ref.~\citenum{Kang1997} and describe interaction with electromagnetic field in the form of Eq.~(\ref{V1}). We show that this results in a consistent description and proceed to calculate the exciton radiative lifetime, $\tau_0$, in a single valley. We will use the gauge where $\varphi({\bf r})=0$.

When the intervalley coupling is neglected, the exciton spin degeneracy is lifted by the electron-hole exchange interaction leading to a splitting between the
exciton bright and dark states which remain valley-degenerate. Recently it has been shown~\cite{Avdeev2020} that
the inter-valley electron-hole exchange interaction leads to a formation of the valley-symmetric ultrabright spin-triplet state of the direct exciton (i.e. exciton with the electron and the hole sharing a valley) and renders all other exciton states optically inactive. The splitting between 
the ultrabright triplet and all the other states is $N=4$ times as large as the spin splitting between the bright and dark states obtained in the model of independent valleys. 
Mixing of the exciton states from different valleys at the surface of the QD
distorts the ultrabright state~\cite{Avdeev2020}. Nevertheless, physics underlying its formation needs special attention.

There are different ways to describe excitons in semiconductors~\cite{Agranovich,Gupalov1998,Cho1999,goupalov2003a,goupalov2003}. Exciton with only direct electron-hole Coulomb interaction taken into account is known as {\it mechanical} exciton~\cite{Agranovich}. When electron-hole exchange interaction is included, the resulting two-particle excitation is called the {\it Coulomb} exciton~\cite{Agranovich}. If the short-range part of the electron-hole exchange is neglected, then interaction of the Coulomb exciton with the transverse electro-magnetic field of light is equivalent to interaction of the mechanical exciton with the full Maxwell field including the longitudinal long-wavelength electric field induced by the macroscopic polarization~\cite{Agranovich,Gupalov1998,Cho1999,goupalov2003a}. In Ref.~\citenum{Avdeev2020} emergence of the ultrabright state was described for a Coulomb exciton. Now the question arises, how
formation of the ultrabright state can be described in terms of the mechanical exciton interacting with the longitudinal electric field.
The advantage of this treatment is that the interaction with the transverse field of light, associated with radiation effects, can be included in a natural way~\cite{Gupalov1998,goupalov2003}. Thus, this approach will be used throughout our paper. In the second part of the present work we will show that the valley coherence resulting in formation of the ultrabright state is akin to superradiance and that the radiative lifetime $\tau_r$ of the ultra-bright state is related to the radiative lifetime $\tau_0$ calculated taking into account only one valley as $\tau_r=\tau_0/N$.

We conclude our work by taking into account effects of the valley mixing and show that resulting exciton radiative lifetimes in PbX QDs are in the microsecond range in agreement with experimental studies on the photoluminescence decay.

\section{Electronic states in a single valley}\label{ehground}
Bulk PbX compounds are direct-gap semiconductors with rocksalt crystal structure and band extrema in the four inequivalent $L$-points of the Brillouin zone. Neglecting the valley anisotropy, conduction- and valence-band electronic states near $L$-points are described by the Hamiltonian~\cite{Kang1997} 
\begin{equation}
\label{dimmock}
\hat{H} =
\left[
\begin{matrix}
\frac{E_g}{2} - \alpha_c \, \Delta 
&
-{\rm i} \, \hbar v_0 \left( {\bm \sigma} {\bm \nabla} \right)
\cr
-{\rm i} \, \hbar v_0 \left( {\bm \sigma} {\bm \nabla} \right)
&
- \frac{E_g}{2} + \alpha_v \, \Delta 
\cr
\end{matrix}
\right] \,,
\end{equation}
where $E_g$ is the energy gap, $\sigma_{\beta}$ ($\beta=x,y,z$) are the 2$\times$2 Pauli matrices, $v_0$ is the Fermi velocity in the gapless limit, the coefficients $\alpha_c$ and $\alpha_v$ stem from the contributions of the remote bands to the conduction and valence bands energy dispersion, and $\Delta$ is the three-dimensional Laplace operator. Formally, the Hamiltonian (\ref{dimmock}) differs from the Dirac Hamiltonian by the diagonal terms. 
The present work reveals the role of these terms in the optical properties of quantum dots made of narrow-gap semiconductors.

The Schr\"odinger equation for the single-particle quantum states in a spherical QD of the radius $R$ has the form
\begin{equation}
\hat{H} \, \left[
\begin{matrix}
\hat{u}({\bf r}) \cr
\hat{v}({\bf r})
\end{matrix}
\right]=E \, \left[
\begin{matrix}
\hat{u}({\bf r}) \cr
\hat{v}({\bf r})
\end{matrix}
\right] \,,
\label{schroed}
\end{equation}
with the boundary conditions
\begin{equation}
\hat{u}(|{\bf r}| = R) = \hat{v}(|{\bf r}| = R) = 0\:.
\end{equation}
Here $\hat{u}({\bf r})$ and $\hat{v}({\bf r})$ are conduction- and valence-band spinor components forming a bispinor envelope function. 
It is convenient to rewrite Eq.~({\ref{schroed}}) as an equivalent set of two equations for the spinor envelopes
\begin{subequations}
\begin{align} \label{Sch1}
&\left( \frac{E_g}{2} - E - \alpha_c \, \Delta \right) \hat{u}({\bf r})  - {\rm i} \hbar v_0 \left( {\bm \sigma} {\bm \nabla} \right) \hat{v}({\bf r}) = 0 \:, 
\\ 
-& {\rm i}  \hbar v_0 \left( {\bm \sigma} {\bm \nabla} \right) \hat{u}({\bf r}) + \left(- \frac{E_g}{2} - E + \alpha_v \, \Delta \right) \hat{v}({\bf r}) = 0\:. \label{Sch2}
\end{align}
\end{subequations}

Electronic states in a spherically symmetric system can be characterized by the total angular momentum $F$, its projection $M$ onto an arbitrary axis, and parity. The ground state of the conduction-band electron confined in a spherical PbX QD has the total angular momentum $F_c=1/2$ and the odd parity~\cite{Kang1997}. The corresponding solution of Eq.~(\ref{schroed}) can be constructed as follows. We first look for a solution of Eq.~(\ref{schroed}) in the form
\begin{subequations}\label{sol1}
\begin{equation}
\hat{u}({\bf r})=A \, j_{F_c-1/2}(kr) \, \hat{\Omega}^{F_c-1/2}_{F_c,M_c}
\left( \frac{\bf r}{r} \right) \,,
\label{sol11}
\end{equation}
\begin{equation}
\hat{v}({\bf r})=B \, j_{F_c+1/2}(kr) \, \hat{\Omega}^{F_c+1/2}_{F_c,M_c}
\left( \frac{\bf r}{r} \right) \,,
\label{sol12}
\end{equation}
\end{subequations}
where $\hat{\Omega}^{F_c \pm 1/2}_{F_c,M_c}$ is the spherical spinor~\cite{Varshalovich} and $j_{F_c \pm 1/2}(kr)$ is the spherical Bessel function.
Using 
\begin{equation}
\left( {\bm \sigma} {\bm \nabla} \right) j_{F_c \pm 1/2} (kr) \, \hat{\Omega}^{F_c \pm 1/2}_{F_c,M_c}=
\mp k \, j_{F_c \mp 1/2} (kr) \, \hat{\Omega}^{F_c \mp 1/2}_{F_c,M_c}
\label{identity_j}
\end{equation}
we obtain from Eq.~(\ref{Sch1}) 
\[
B=i \, \frac{E_g+2 \, \alpha_c \, k^2 - 2 \, E}{2 \, \hbar v_0 \, k} \, A \equiv i \, \rho(k) \, A \,,
\]
while Eq.~(\ref{Sch2}) yields
\begin{equation}
k=\sqrt{\Pi+\Sigma} \,, 
\label{k}
\end{equation}
where
\[
\Sigma=\frac{E \, (\alpha_v-\alpha_c) - \hbar^2 v_0^2 -E_g \, (\alpha_v+\alpha_c)/2}{2 \alpha_c \alpha_v} \,,
\]
\[
\Pi=\sqrt{\Sigma^2 + \frac{E^2 - (E_g/2)^2}{\alpha_c \alpha_v}} \,.
\]
Another solution of the bispinor equation~(\ref{schroed})
is given by
\begin{subequations}\label{sol2}
\begin{eqnarray}
&&\hat{u}({\bf r})=C \, i^{(1)}_{F_c-1/2}(\kappa r) \, \hat{\Omega}^{F_c-1/2}_{F_c,M_c}
\left( \frac{\bf r}{r} \right) \,, \label{sol21}\\
&&\hat{v}({\bf r})=D \, i^{(1)}_{F_c+1/2}(\kappa r) \, \hat{\Omega}^{F_c+1/2}_{F_c,M_c}
\left( \frac{\bf r}{r} \right) \,, \label{sol22}
\end{eqnarray}
\end{subequations}
where $i^{(1)}_{F_c \pm 1/2}(\kappa r)$ is the modified spherical Bessel function.
Using 
\begin{equation}
\left( {\bm \sigma} {\bm \nabla} \right) i^{(1)}_{F_c \pm 1/2} (\kappa r) \, \hat{\Omega}^{F_c \pm 1/2}_{F_c,M_c}=
- \kappa \, i^{(1)}_{F_c \mp 1/2} (\kappa r) \, \hat{\Omega}^{F_c \mp 1/2}_{F_c,M_c}
\label{identity_i}
\end{equation}
we obtain from Eq.~(\ref{Sch1})
\[
D={\rm i} \, \frac{E_g-2 \, \alpha_c \, \kappa^2 - 2 \, E}{2 \, \hbar v_0 \, \kappa} \, C \equiv {\rm i} \, \mu(\kappa) \, C\,,
\]
while Eq.~(\ref{Sch2}) yields
\begin{equation}
\kappa=\sqrt{\Pi-\Sigma} \,.
\label{kappa} 
\end{equation}
From the condition that a linear combination of the solutions~(\ref{sol1}) and~(\ref{sol2}) vanishes at $r=R$,
we obtain the dispersion equation for $k \equiv k_c$, $\kappa \equiv \kappa_c$~\cite{Kang1997}
\begin{multline}
i^{(1)}_{F+1/2}(\kappa_c R) \,
j_{F-1/2}(k_c R) \,
\mu(\kappa_c) \\ 
-i^{(1)}_{F-1/2}(\kappa_c R) \,
j_{F+1/2}(k_c R) \,
\rho(k_c)=0 \,,
\end{multline}
which yields the energy of the confined conduction-band electron state ($E>0$).
The radial wave functions are~\cite{Kang1997}
\begin{widetext}
\begin{subequations}\label{zc}
\begin{eqnarray}
z^c_{F_c-1/2}(r)&=&A_c \left[ j_{F_c-1/2}(k_c r) -\frac{j_{F_c-1/2}(k_c R)}{i^{(1)}_{F_c-1/2}(\kappa_c R)}
\, i^{(1)}_{F_c-1/2}(\kappa_c r) \right] \,,
\label{z0c}\\
z^c_{F_c+1/2}(r)&=&A_c \left[ \rho(k_c) \, j_{F_c+1/2}(k_c r) - \mu(\kappa_c) \, \frac{j_{F_c-1/2}(k_c R)}{i^{(1)}_{F_c-1/2}(\kappa_c R)}
\, i^{(1)}_{F_c+1/2}(\kappa_c r) \right] \,,
\label{z1c}
\end{eqnarray}
\end{subequations}
\end{widetext}
where $A_c$ is a normalization constant determined by the condition
\begin{equation}
\int\limits_0^R dr r^2 \left[ z_{F_c-1/2}^{c \, 2} (r)
+  z_{F_c+1/2}^{c \, 2} (r) \right]=1 \,.
\end{equation}
In this work we will need these functions only for $F_c=1/2$. Thus, for the ground state of the conduction-band electron confined in a spherical PbX QD
we have a bispinor wave function
\begin{subequations}\label{uvce}
\begin{eqnarray}
&& \hat{u}^c_{1/2,M_c}({\bf r})
=z_0^c(r) \, \hat{\Omega}^0_{1/2,M_c}
\left( \frac{\bf r}{r} \right) \,,
\label{uce} \\
&& \hat{v}^c_{1/2,M_c}({\bf r})
={\rm i} \,  z_1^c(r) \, \hat{\Omega}^1_{1/2,M_c}
\left( \frac{\bf r}{r} \right) \,.
\label{vce}
\end{eqnarray}
\end{subequations}

Similarly to the conduction electrons, the ground state of the valence-band hole confined in a spherical PbX QD has the total angular momentum $F_h=1/2$ and the even parity. We will again construct two solutions of Eq.~(\ref{schroed}) in a free space and impose a boundary condition on their linear combination. This time for the first solution of Eq.~(\ref{schroed}) we use the substitution
\begin{subequations}
\begin{eqnarray}
&&\hat{u}({\bf r})=A \, j_{F_h+1/2}(kr) \, \hat{\Omega}^{F_h+1/2}_{F_h,M_h}
\left( \frac{\bf r}{r} \right) \,,  \\
&&\hat{v}({\bf r})=B \, j_{F_h-1/2}(kr) \, \hat{\Omega}^{F_h-1/2}_{F_h,M_h}
\left( \frac{\bf r}{r} \right)  
\end{eqnarray}
\end{subequations}
and get
\begin{equation}
B=-{\rm i} \, \rho(k) \, A \,.
\end{equation}
For the second solution of Eq.~(\ref{schroed}) we try
\begin{subequations}
\begin{eqnarray}
&&\hat{u}({\bf r})=C \, i^{(1)}_{F_h+1/2}(\kappa r) \, \hat{\Omega}^{F_h+1/2}_{F_h,M_h}
\left( \frac{\bf r}{r} \right) \,,
\\
&&\hat{v}({\bf r})=D \, i^{(1)}_{F_h-1/2}(\kappa r) \, \hat{\Omega}^{F_h-1/2}_{F_h,M_h}
\left( \frac{\bf r}{r} \right) \,,
\end{eqnarray}
\end{subequations}
and we again obtain
\begin{equation}
D={\rm i} \, \mu(\kappa) \, C \,.
\end{equation}
From the condition that a linear combination of these two solutions vanishes at $r=R$
we obtain the dispersion equation for $k=k_v$ and $\kappa=\kappa_v$~\cite{Kang1997} 
\begin{multline}
i^{(1)}_{F_h-1/2}(\kappa_v R) \,
j_{F_h+1/2}(k_v R) \,
\mu(\kappa_v) \\
+i^{(1)}_{F_h+1/2}(\kappa_v R) \,
j_{F-1/2}(k_v R) \,
\rho(k_v)=0 \,,
\end{multline}
which yields the energy of the confined valence-band hole state ($E<0$).
The radial wave functions are~\cite{Kang1997}
\begin{widetext}
\begin{subequations}\label{zv}
\begin{eqnarray}
z^v_{F_h+1/2}(r)&=&B_v \, \left[j_{F_h+1/2}(k_v r) -\frac{j_{F_h+1/2}(k_v R)}{i^{(1)}_{F_h+1/2}(\kappa_v R)}
\, i^{(1)}_{F_h+1/2}(\kappa_v r) \right] \,, \label{z1v}\\
z^v_{F_h-1/2}(r)&=&B_v \, \left[ \rho(k_v) \, j_{F_h-1/2}(k_v r) + \mu(\kappa_v) \, \frac{j_{F_h+1/2}(k_v R)}{i^{(1)}_{F_h+1/2}(\kappa_v R)}
\, i^{(1)}_{F_h-1/2}(\kappa_v r) \right] \,, \label{z0v}
\end{eqnarray}
\end{subequations}
\end{widetext}
where $B_v$ is a normalization constant determined by the condition
\begin{equation}
\int\limits_0^R dr r^2 \left[ z_{F_h-1/2}^{v \, 2} (r)
+  z_{F_h+1/2}^{v \, 2} (r) \right]=1 \,.
\end{equation}
The resulting bispinor wave function for the ground state of the valence-band hole confined in a spherical PbX QD
takes the form 
\begin{subequations}\label{uvvh}
\begin{eqnarray}
&&\hat{u}^v_{1/2,M_h}({\bf r})=z_1^v(r) \, \hat{\Omega}^1_{1/2,M_h}
\left( \frac{\bf r}{r} \right) \,,
\label{uvh}\\ &&
\hat{v}^v_{1/2,M_h}({\bf r})=- {\rm i}  z_0^v(r) \, \hat{\Omega}^0_{1/2,M_h}
\left( \frac{\bf r}{r} \right) \,. \label{vvh}
\end{eqnarray}
\end{subequations}
The corresponding valence-band electron states can be obtained applying the time-reversal operator:
\begin{subequations}\label{uvve}
\begin{eqnarray}
&&K \hat{u}^v_{1/2,M_h}
=(-1)^{3/2-M_h} \, {\rm i} z_1^v(r) \, \hat{\Omega}^1_{1/2,-M_h} \left( \frac{\bf r}{r} \right)\,,
\label{uve} \\ 
&&K \hat{v}^v_{1/2,M_h}
=(-1)^{3/2-M_h} \, z_0^v(r) \, 
\hat{\Omega}^0_{1/2,-M_h} \left( \frac{\bf r}{r} \right) \,. \label{vve}
\end{eqnarray}
\end{subequations}

\section{Optical excitation of an exciton in a single valley} \label{formalism}

Neglecting the electron-hole exchange interaction,
the ground exciton level in a given valley of a PbX QD is four-fold spin-degenerate. The four exciton states can be labeled by the total exciton angular momentum ${\cal F}$ and its projection ${\cal F}_z$ onto the $z$ axis which, in the isotropic case, can be chosen arbitrarily. The optically active states have ${\cal F} = 1$.
Using Wigner $3jm$-symbols~\cite{Varshalovich} these states can be written as
\begin{multline}
|X,1 {\cal F}_z \rangle=(-1)^{{\cal F}_z} \, \sqrt{3} \\\times \sum\limits_{M_c,M_h} \left(
\begin{smallmatrix}
\frac{1}{2}&\frac{1}{2}&1\cr
M_c&M_h&-{\cal F}_z
\end{smallmatrix}
\right) |c, M_c \rangle |v, M_h \rangle \,,
\label{exc}
\end{multline}
where $|c, M_c \rangle$ and $|v, M_h \rangle$ refer to the states of the conduction-band electron and valence-band hole whose bispinor wave functions are given by Eqs.~(\ref{uvce}) and~(\ref{uvvh}), respectively.

In the linear optical regime, the state of the optically excited QD can be represented as its ground state $| 0 \rangle$ and a small correction:
\begin{equation}
|t \rangle =|0 \rangle +C_{{\cal F}_z}(t)  | X, 1 {\cal F}_z \rangle  {\rm e}^{-{\rm i} \omega_0 t} \,,
\label{t}
\end{equation}
where $\omega_0$ is the resonance frequency of the mechanical exciton.
The coefficient $C_{{\cal F}_z}(t)$ can be found from the Schr\"odinger equation
\begin{equation}
{\rm i} \hbar \frac{\partial}{\partial t} |t \rangle= (\hat{H}+\hat{V}(t)) |t \rangle \,,
\label{schroed2}
\end{equation}
where $\hat{H}$ is the unperturbed two-particle Hamiltonian describing the electron-hole pair and
\begin{multline}
\label{V}
\hat{V}(t)=-\frac{1}{c} \int  {\bf j}({\bf r}) \, {\bf A}({\bf r},t) d {\bf r} \\
=\frac{\rm i}{\omega} \int \sum\limits_{\mu} j_{\mu}({\bf r}) \, \frac1{V} \sum\limits_{\bf q} E^{\mu}({\bf q},t) \, e^{{\rm i} {\bf qr}} d {\bf r} \\
=\frac{\rm i}{\omega V} \sum\limits_{\mu {\bf q}}  j_{\mu}(-{\bf q}) E^{\mu}({\bf q},t) 
\end{multline}
is the perturbation describing the light-matter interaction with the plane electromagnetic wave characterized by the vector potential ${\bf A}({\bf r},t)$, electric field ${\bf E}({\bf r},t)=V^{-1} \sum\limits_{{\bf q}} {\bf E}({\bf q},t) \, e^{i{\bf qr} }$, and frequency $\omega$,
$V$ is the normalization volume, ${\bf j}({\bf r})$ is the current density operator, and we distinguish between co- and contra-variant cyclic components of vectors.
In writing Eq.~(\ref{V}) we assumed ${\bf A}({\bf r},t) \propto e^{-i \omega t}$ and neglected the complex conjugated term leading to a non-resonant contribution to the polarization.
Substituting~Eq. (\ref{t}) into Eq.~(\ref{schroed2}) and multiplying by $\langle X, 1 {\cal F}_z|$ from the left we obtain
\begin{equation}
\hbar \frac{\partial C_{{\cal F}_z}}{\partial t}=
\frac{e^{- {\rm i}(\omega-\omega_0)t}}{\omega} \, \Lambda
\,,
\label{C1}
\end{equation}
where
\begin{equation}
\Lambda= \frac{1}{V}\sum\limits_{\mu {\bf q}} \langle X, 1 {\cal F}_z | j_{\mu}(-{\bf q})|0 \rangle \, E^{\mu}({\bf q}) \,.
\end{equation}
Integrating Eq.~(\ref{C1}) we get
\begin{equation}
C_{{\cal F}_z}(t)=\frac{{\rm i}\, \Lambda \, e^{- {\rm i}(\omega-\omega_0)t} }{\hbar \omega (\omega-\omega_0+{\rm i}0)} 
\,.
\label{C2}
\end{equation}
Then for the Fourier-component of polarization we obtain
\begin{equation}
P_{{\rm exc}}^{\sigma}({\bf q},\omega)=\frac{\rm i}{\omega} \langle t | j^{\sigma}({\bf q})|t \rangle_{\omega}=-
\frac{\langle 0|j^{\sigma}({\bf q}) |X, 1 {\cal F}_z \rangle}{\hbar \omega^2 (\omega-\omega_0+{\rm i} 0)}  \, \Lambda \,.
\label{polar1}
\end{equation}
Thus, the linear susceptibility of the QD has a tensor character. It would become scalar only if $\langle 0|j^{\sigma}({\bf q}) |X, 1 {\cal F}_z \rangle
\propto \delta_{\sigma,{\cal F}_z}$.

From Maxwell equations
\begin{subequations}
\begin{align}
  \left[ {\bm \nabla} \times \left[ {\bm \nabla} \times {\bm E}({\bm r}) \right] \right] & = k_0^2 {\bm D} ({\bm r})\,, \\
  {\bm \nabla} \cdot {\bm D} ({\bm r}) = {\bm \nabla} ( \varepsilon_b {\bm E}({\bm r}) & + 4 \pi {\bm P}_{{\rm exc}} ) =0\,,
\end{align}
\end{subequations}
where $k_0=\omega/c$ and $\varepsilon_b$ is the background permittivity, we have
\begin{multline}
(-q^2+k^2) \, E^{\mu}({\bf q})=-k_0^2 \, 4 \pi P_{{\rm exc}}^{\mu}({\bf q},\omega)
\\+\frac{1}{\varepsilon_b} q^{\mu} \sum\limits_{\sigma} q_{\sigma} P_{{\rm exc}}^{\sigma}({\bf q},\omega) \, ,
\label{Maxwell}
\end{multline}
where $k=\sqrt{\varepsilon_b} \, k_0$. Equation~(\ref{Maxwell}) yields 
\begin{widetext}
\begin{eqnarray}
E^{\mu}({\bf q})=E^{\mu (0)} \delta_{{\bf q},{\bf k}}&-&\frac{k_0^2}{q^2-k^2} \frac{4 \pi}{\hbar \omega^2 (\omega-\omega_0+{\rm i}0)}
\langle 0|j^{\mu}({\bf q}) |X, 1 {\cal F}_z \rangle \Lambda  \\
 &+&\ \frac{1}{\varepsilon_b} \frac{q^{\mu}}{q^2-k^2} \frac{4 \pi}{\hbar \omega^2 (\omega-\omega_0+{\rm i}0)} \Lambda \sum\limits_{\sigma} q_{\sigma} \langle 0|j^{\sigma}({\bf q}) |X, 1 {\cal F}_z \rangle \,. \nonumber
\end{eqnarray}
\end{widetext}
Multiplying this equation by $\langle X, 1 {\cal F}_z | j_{\mu}(-{\bf q})|0 \rangle$ and summing over ${\bf q}$ and $\mu$ we arrive at
\begin{equation}\label{lambda1_def}
\Lambda=\Lambda^{(0)}+ \Lambda \frac{\Xi}{\omega-\omega_0+{\rm i}0} \,,
\end{equation}
or
\begin{equation}
\Lambda= \frac{\Lambda^{(0)}}{ \omega-\omega_0 - \Xi} \,,
\label{lambda1}
\end{equation}
where 
\begin{subequations}\label{LXi}
\begin{eqnarray}
&\Lambda^{(0)} = \sum\limits_{\mu} \langle X, 1 {\cal F}_z | j_{\mu}(-{\bf k})|0 \rangle E^{\mu (0)}\:, \\
&\Xi=\frac{4 \pi}{\varepsilon_b \hbar \omega^2 V} \sum\limits_{\bf q} \sum\limits_{\mu,\sigma} \frac{q^{\mu} q_{\sigma}-k^2 \delta_{\mu,\sigma}}{q^2-k^2-{\rm i}0}
F_{\mu}^{\sigma}({\bf q}) \,,
\label{Xi} \\
&F_{\mu}^{\sigma}({\bf q}) =\langle X, 1 {\cal F}_z | j_{\mu}(-{\bf q})|0 \rangle \langle 0|j^{\sigma}({\bf q}) |X, 1 {\cal F}_z \rangle\,\nonumber.
\end{eqnarray}
\end{subequations}
One can see from Eq.~(\ref{lambda1}) that the real and imaginary parts of $\Xi$ determine, respectively, the resonant frequency renormalization $\delta \omega$ due to the electron-hole long-range exchange interaction and the radiative lifetime $\tau_0$ as follows
\begin{equation}
\Xi=\delta \omega -\frac{\rm i}{2 \tau_0} \,.
\label{Xideltaomega}
\end{equation}

\section{Interband matrix elements of coordinate, velocity, and current}

The current density operator for a particle at the point ${\bf r}$ is defined as
\begin{equation}
{\bf j}({\bf r})=\frac{e}{2} \, [ {\bf v}_e \delta({\bf r}-{\bf r}_e) + \delta({\bf r}-{\bf r}_e) {\bf v}_e ]\,,
\end{equation}
where ${\bf v}_e$ is the velocity operator
\begin{equation} \label{velocity}
{\bf v}_e = \frac{\rm i}{\hbar} (\hat{H} {\bf r}_e - {\bf r}_e \hat{H})\:.
\end{equation}
From this definition we get an expression for the Fourier transform of the current entering Eq.~(\ref{Xi}):
\begin{equation}
{\bf j}({\bf q})=\frac{e}{2} \, ( {\bf v}_e {\rm e}^{-i {\bf qr}_e} + {\rm e}^{-i {\bf qr}_e} {\bf v}_e ) \,.
\label{jqdef}
\end{equation}

First we discuss the longitudinal current component parallel to ${\bf q}$. 
For the Fourier transform of ${\bm \nabla}{\bf j}$ we get from the continuity equation $\bm{\nabla} {\bf j}({\bf r}) = -\frac{\partial}{\partial t} \rho({\bf r})$
\begin{equation}\label{eq:qj}
{\bf q j}({\bf q})={\rm i} e \frac{d}{dt} \left( {\rm e}^{-{\rm i} {\bf q r}_e}\right)=-\frac{e}{\hbar}  \left[H,{\rm e}^{- {\rm i} {\bf q r}_e} \right] \,.
\end{equation}
For its matrix elements we have
\begin{multline}
\langle f| {\bf q j}({\bf q}) |i \rangle=-e  \frac{E_f-E_i}{\hbar}  \langle f| {\rm e}^{- {\rm i} {\bf  q r}_e} | i \rangle
\\=- e \, \omega \, \langle f| {\rm e}^{- {\rm i} {\bf  q r}_e} | i \rangle \,,
\label{qj_matrel}
\end{multline}
where $|i \rangle$ and $|f \rangle$ stay for the initial and final states, respectively. 
Replacing ${\bf q}$ by $- {\bf q}$ in Eq.~(\ref{qj_matrel})  we find
\begin{equation}
\langle f | {\bf q j}({-\bf q}) | i \rangle=
e \, \omega \, \langle f| {\rm e}^{{\rm i} {\bf  q r}_e} | i \rangle \,.
\label{mqj_matrel}
\end{equation}
Using for the final and initial states the bispinors of Eqs.~(\ref{uvce}) and~(\ref{uvve}), respectively,
we obtain
\begin{multline}
\langle
c, M_c|{\rm e}^{{\rm i} {\bf qr}_e} |v, K M_h \rangle={\cal K}(q) \, \sqrt{8 \pi} \, Y^*_{1 \, M_c+M_h} \left( \frac{\bf q}{q} \right) \,  \\\times (-1)^{M_c+M_h} \,
\left(\begin{smallmatrix}
\frac{1}{2}&\frac{1}{2}&1\cr
-M_h&-M_c&M_c+M_h
\end{smallmatrix}\right) \,,
\end{multline}
where
\begin{equation}
{\cal K}(q)=\int\limits_0^R dr r^2 \left[ z_0^c(r) z_1^v(r) -
z_0^v(r) z_1^c(r) \right] \, j_1(qr) \,.
\end{equation}
It follows From Eq.~(\ref{exc}) that
\begin{equation}
\langle X, 1 {\cal F}_z |{\rm e}^{{\rm i} {\bf qr}_e} |0 \rangle=\sqrt{\frac{8 \pi}{3}} \, {\cal K}(q) \, Y^*_{1 \, {\cal F}_z} \left( \frac{\bf q}{q} \right)  \,.
\end{equation}   
Then from Eqs.~(\ref{qj_matrel}),~(\ref{mqj_matrel}) one obtains
\begin{subequations}\label{qj}
\begin{equation}
\langle 0 | {\bf q j}({\bf q}) | X, 1 {\cal F}_z \rangle
=e \, \omega \,  \sqrt{\frac{8 \pi}{3}} \, {\cal K}(q) \, Y_{1 \, {\cal F}_z} \left( \frac{\bf q}{q} \right) \,,
\label{qj1}
\end{equation}   
\begin{equation}
\langle X, 1 {\cal F}_z | {\bf q j}(-{\bf q}) | 0 \rangle
=e \, \omega \,  \sqrt{\frac{8 \pi}{3}} \, {\cal K}(q) \, Y^*_{1 \, {\cal F}_z} \left( \frac{\bf q}{q} \right) \,.
\label{qj2}
\end{equation}
\end{subequations}

Now we will proceed to calculate the matrix elements of the ${\bf j}({\bf q})$ operator. From Eqs.~(\ref{dimmock}),~(\ref{jqdef}) and the definition of the velocity operator (\ref{velocity}) we get an explicit form of ${\bf j}({{\bf q}})$:
\begin{widetext}
\begin{equation}
{\bf j}({{\bf q}})= \frac{e}{\hbar}
\left[
\begin{matrix}
- \alpha_c {\bf q} {\rm e}^{-{\rm i} {\bf qr}_e} - 2 {\rm i} \alpha_c {\rm e}^{-{\rm i} {\bf qr}_e} \bm{\nabla}_e & \hbar v_0 \bm{\sigma} {\rm e}^{-{\rm i} {\bf qr}_e} \\
\hbar v_0 \bm{\sigma} {\rm e}^{-{\rm i} {\bf qr}_e}&
\alpha_v {\bf q} {\rm e}^{-{\rm i} {\bf qr}_e} + 2 {\rm i} \alpha_v {\rm e}^{-{\rm i} {\bf qr}_e} \bm{\nabla}_e
\end{matrix}
\right] \,,
\label{full}
\end{equation}
where $\bm{\nabla}_e = \partial / \partial {\bf r}_e$.

Its matrix element has the form
\begin{equation}
\langle 0 |j_{\beta}({\bf q})|X, 1 {\cal F}_z \rangle=a(q) \, (-1)^{{\cal F}_z} \, \delta_{\beta,-{\cal F}_z}
+ b(q) \, (-1)^{{\cal F}_z+\beta} \left(
\begin{matrix}
1&1&2\\
\beta & {\cal F}_z & -\beta-{\cal F}_z
\end{matrix} 
\right) \, Y_{2 \, \beta+{\cal F}_z} \left( \frac{{\bf q}}{q} \right) \,,
\label{jqmatrelem}
\end{equation}
where 
\begin{multline}
a(q)=e \, \sqrt{2} \left\{ v_0 \, \int\limits_0^R dr \, r^2 \left[ z_0^c(r) \, z_0^v(r)-\frac{1}{3} \, z_1^c(r) \, z_1^v(r) \right] \, j_0(qr)
\right.
\\ 
\left.
-\ \frac{q}{3 \hbar} \int\limits_0^R dr \, r^2 \left[\alpha_c \, z_0^c(r) \, z_1^v(r)- \alpha_v \, z_0^v(r) \, z_1^c(r) \right] \, j_1(qr)
\right. 
\\ 
\left.
+\ \frac{2}{3 \hbar} \, \int\limits_0^R dr \, r^2 \left[\alpha_c \, \frac{\partial z_0^c(r)}{\partial r} \, z_1^v(r)- \alpha_v \, \frac{\partial z_0^v(r)}{\partial r} \, z_1^c(r) \right] \, j_0(qr) \right\} \,,
\end{multline}
\begin{multline}
b(q) =4 \,e \, \sqrt{\pi} \left\{ - \frac{v_0}{\sqrt{15}} \, \int\limits_0^R dr \, r^2 z_1^c(r) \, z_1^v(r) \, j_2(qr)
\right. 
\\ 
\left.
-\ \frac{q}{3 \hbar} \int\limits_0^R dr \, r^2 \left[\alpha_c \, z_0^c(r) \, z_1^v(r)- \alpha_v \, z_0^v(r) \, z_1^c(r) \right] \, j_1(qr)
\right. 
\\ 
\left.
- \frac{2}{3 \hbar} \, \int\limits_0^R dr \, r^2 \left[\alpha_c \, \frac{\partial z_0^c(r)}{\partial r} \, z_1^v(r)- \alpha_v \, \frac{\partial z_0^v(r)}{\partial r} \, z_1^c(r) \right] \, j_2(qr)
\right\} \,. 
\end{multline}
If only the first term were present in the right-hand side of Eq.~(\ref{jqmatrelem}), then the linear susceptibility of the QD would be scalar. If we now calculate matrix elements of ${\bf q j}({\bf q})$ using Eq.~(\ref{jqmatrelem}) and compare it to Eq.~(\ref{qj1}) we come to
\begin{multline}
\label{bigeq}
\hbar \, \omega \, {\cal K}(q)=
-q^2 \, \int\limits_0^R dr \, r^2 \left[\alpha_c \, z_0^c(r) \, z_1^v(r)- \alpha_v \, z_0^v(r) \, z_1^c(r) \right] \, j_1(qr)
\\
+2 \, \int\limits_0^R dr \, r^2 \left[\alpha_c \, \frac{\partial z_0^c(r)}{\partial r} \, z_1^v(r)- \alpha_v \, \frac{\partial z_0^v(r)}{\partial r} \, z_1^c(r) \right] \, \frac{d j_1(qr)}{dr}
\\
+v_0 \, \hbar \, q \, \int\limits_0^R dr \, r^2 \left[ z_0^c(r) \, z_0^v(r)-\frac{1}{3} \, z_1^c(r) \, z_1^v(r) \right] \, j_0(qr)
\\
+v_0 \, \hbar \, q \, \frac{4}{3} \, \int\limits_0^R dr \, r^2 \, z_1^c(r) \, z_1^v(r) \, j_2(qr) \,.
\end{multline}
\end{widetext}
In the lowest order in $q$ we have $j_0(qr) \approx 1, j_1(qr) \approx qr/3$ and 
Eq.~(\ref{bigeq}) yields
\begin{equation}\label{therelation}
\frac{\hbar \, \omega}{3} R I_r 
= \frac{2}{3} \frac1R I_{\alpha}\, +v_0 \, \hbar \, I_p \,,
\end{equation}
where we defined the following integrals:
\begin{subequations}
\begin{equation}
I_r = \frac1{R} \int\limits_0^R dr \, r^3 \left[ z_0^c(r) \, z_1^v(r)-z_0^v(r) \, z_1^c(r) \right]\,,
\end{equation}
\begin{equation}
I_{\alpha} = R \int\limits_0^R dr \, r^2 \left[\alpha_c \, \frac{\partial z_0^c(r)}{\partial r} \, z_1^v(r)- \alpha_v \, \frac{\partial z_0^v(r)}{\partial r} \, z_1^c(r) \right]\,,
\end{equation}
\begin{equation}
I_p = \int\limits_0^R dr \, r^2 \left[ z_0^c(r) \, z_0^v(r)-\frac{1}{3} \, z_1^c(r) \, z_1^v(r) \right]\,.
\end{equation}
\end{subequations}

\begin{figure}[tb]
\includegraphics[width=0.9\linewidth]{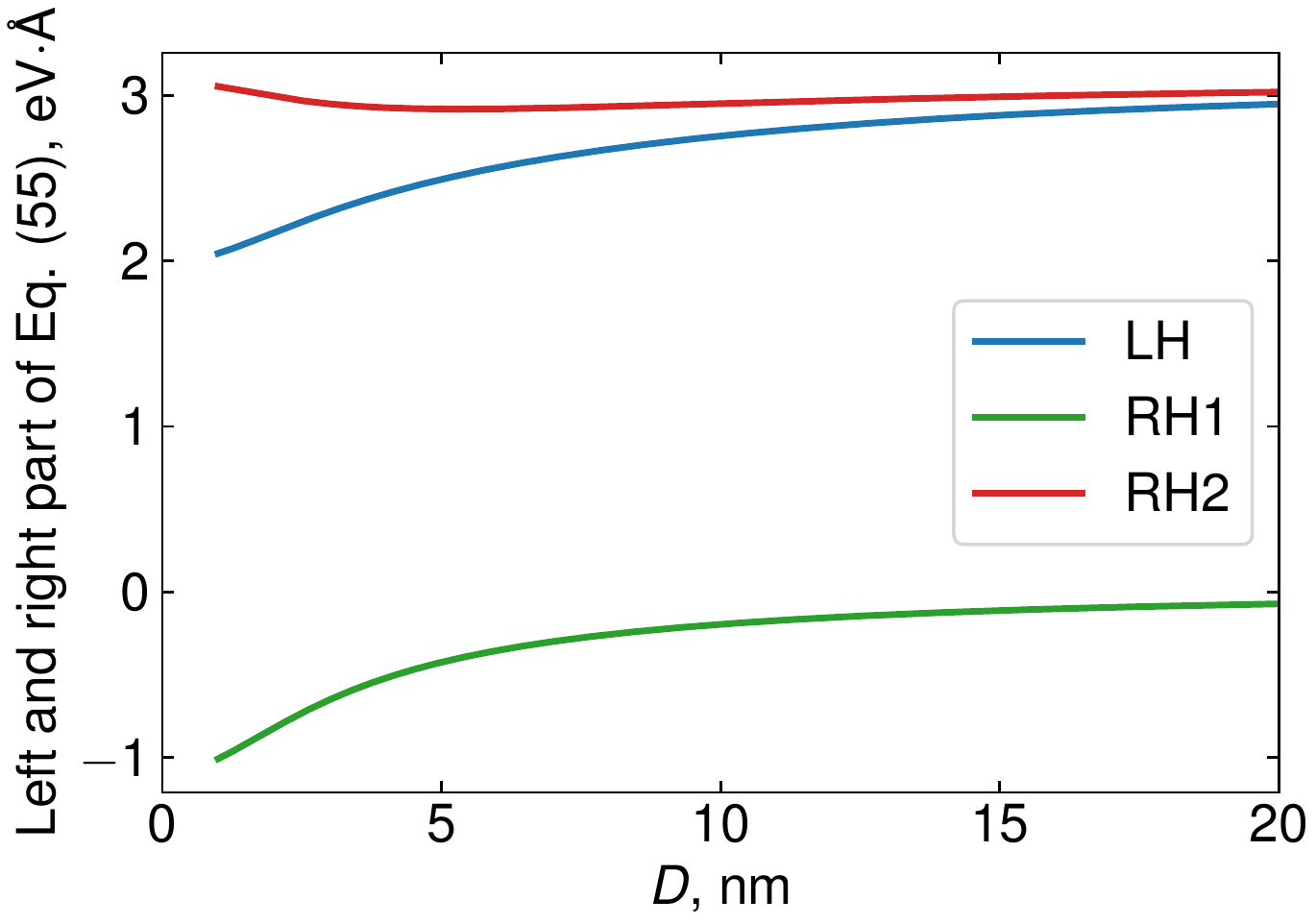}
\caption{
  Size dependences of the terms in the left- and right-hand sides of Eq.~\eqref{therelation} for PbS QDs.
}
\label{fig:therelation}
\end{figure}
In the left-hand side of Eq.~(\ref{therelation}) there is an expression proportional to the interband matrix element of the coordinate
operator while the right-hand side is proportional to the interband matrix element of the velocity operator. The appearance of the first term in the right-hand side of Eq.~(\ref{therelation}) reflects specifics of this problem. Although we derived Eq.~(\ref{therelation}) from the matrix elements of the Fourier transform of $\bm{\nabla}{\bf j}$, it can also be derived from the matrix elements of the velocity operator. The explicit expressions for the interband matrix elements of the coordinate and velocity operators are 
\begin{multline}
\langle c, M_c| \hat{v}_{\beta}| v, KM_h \rangle
\\
= \sqrt{6} \,
(-1)^{1+\beta} \,
\left(
\begin{smallmatrix}
\frac{1}{2}&\frac{1}{2}&1\cr
M_c&M_h&-\beta
\end{smallmatrix}
\right)
\label{A1}
\left\{
v_0 I_p +\frac{2}{3 \hbar} \frac1{R} I_{\alpha}
\right\}
\,,
\end{multline}
\begin{multline}\label{A2} 
\langle c, M_c | r_{\beta}|v,  KM_h \rangle
\\=  i \, \sqrt{\frac{2}{3}} \,
(-1)^{1+\beta} \,
\left(
\begin{smallmatrix}
\frac{1}{2}&\frac{1}{2}&1\cr
M_c&M_h&-\beta
\end{smallmatrix}
\right)
R I_r\,.
\end{multline}

To emphasize the importance of the first term in the right-hand side of Eq.~\eqref{therelation}, in Fig.~\ref{fig:therelation} we show the term in the left-hand side of this equation (LH) and the two terms in the right-hand side (RH1 and RH2) calculated separately as functions of the QD size for the parameters corresponding to PbS \cite{Kang1997}. For the QD diameter about 10~\AA, the absolute value of the term RH1 is about one half of the LH term, it decreases with the increasing QD size. However, its value is still appreciable even for QDs with the diameter of 10~nm. 

\section{Radiative lifetime. Single-valley case}
The matrix elements entering Eq.~(\ref{Xi}) are related to Eq.~(\ref{jqmatrelem}) via
\begin{equation}
\begin{split}\label{vspom}
\langle 0 |j^{\beta}({\bf q})|X, 1 {\cal F}_z \rangle&=(-1)^{\beta} \, \langle 0 |j_{-\beta}({\bf q})|X, 1 {\cal F}_z \rangle \,,
\\
\langle X, 1 {\cal F}_z |j_{\beta}(-{\bf q})|0 &\rangle
\\= (&-1)^{\beta} \, \left[ \langle 0 |j_{-\beta}({\bf -q})|X, 1 {\cal F}_z \rangle \right]^* \,.
\end{split}
\end{equation}
Substituting Eqs.~(\ref{qj}),~(\ref{jqmatrelem}),~(\ref{vspom}) into Eq.~(\ref{LXi}) we obtain
\begin{multline}
\Xi= \frac{4 e^2}{3 \pi \varepsilon_b \hbar } \, \int\limits_0^{\infty} dq \, {\cal K}^2(q) \\ + {\rm i} \frac{k}{\varepsilon_b \hbar} \,  \left\{
\frac23 e^2 {\cal K}^2(k) - \frac{k^2}{\omega^2} \left[ a^2(k) + \frac{b^2(k)}{12 \pi} \right]  \right\}
\,.
\label{Xi1}
\end{multline}
The first term in the right-hand side of Eq.~(\ref{Xi1}) gives the resonant frequency renormalization due to the electron-hole long-range exchange interaction~\cite{Avdeev2020}. For the imaginary part of Eq.~(\ref{Xi1}) we will take the long-wavelength limit:
\begin{multline}
\frac{1}{2 \tau_0}=\frac{2 k^3 \, e^2}{\varepsilon_b \hbar} \, \left\{
- \frac{1}{27} \, \left[ R I_r \right]^2 \right. 
\\ 
 +\ \frac{1}{\omega^2} \left[v_0 I_p \left.
+\frac{2}{3 \hbar} \frac1R I_{\alpha} \right]^2 \right\}.
\end{multline}
Taking into account Eq.~(\ref{therelation}) we finally obtain
\begin{equation}
\frac{1}{\tau_0}=\frac{8}{27} \frac{k^3 e^2}{\varepsilon_b \hbar} R^2 I_r^2\,.
\label{result}
\end{equation}
We note that, in the long-wavelength limit, the radiative lifetime can be obtained using the Fermi golden rule. In this limit
\begin{multline}
\langle i| {\bf j}({\bf q} \rightarrow 0) | f \rangle=\frac{i}{\hbar} \, e \, (E_i-E_f) \, \langle i| {\bf r}_e | f \rangle\\
=-i \, e \, \omega \, \langle i| {\bf r}_e | f \rangle \,,
\end{multline}
in agreement with Eq.~(\ref{result}), cf. Eq.~(\ref{A2}).

\section{Emergence of the ultrabright state}\label{ultrabright}
In this section we will generalize our results for the multi-valley case. When exciton states in different valleys are taken into account,
we need to replace Eq.~(\ref{t}) by
\begin{equation}
|t \rangle =|0 \rangle + \sum\limits_{i=1}^N C_{{\cal F}_z,i}(t) | X, 1 {\cal F}_z, i \rangle \, {\rm e}^{- {\rm i} \omega_i t} \,,
\label{t2}
\end{equation}
where $i$ is the valley index   
and we took into account that only the states of direct excitons, with the electron and the hole from the same valley, can interact with light.
Then, instead of Eqs.~(\ref{polar1}) and~(\ref{lambda1_def}), we get, respectively, 
\begin{equation}
P_{{\rm exc}}^{\sigma}({\bf q},\omega)=
-\sum\limits_{i=1}^N
\frac{\langle 0|j^{\sigma}({\bf q}) |X, 1 {\cal F}_z,i \rangle}{\hbar \omega^2 (\omega-\omega_i+i0)}  \, \Lambda_i 
\label{polar2}
\end{equation}
and
\begin{equation}
\Lambda_j=\Lambda_j^{(0)} -\sum\limits_{i=1}^N \frac{\Xi^{ji}}{\omega_i-\omega-i0} \, \Lambda_i \,,
\label{lambda2}
\end{equation} 
where
\begin{subequations}\label{LXi2}
\begin{eqnarray}
&& \Lambda^{(0)}_j = \sum\limits_{\mu} \langle X, 1 {\cal F}_z,j | j_{\mu}(-{\bf k})|0 \rangle E^{\mu (0)}\:, \hspace{-1 cm}\\
&&  
\Xi^{ji}=\frac{4 \pi}{\varepsilon_b \hbar \omega^2 V} \sum\limits_{\bf q} \sum\limits_{\mu,\sigma} \frac{q^{\mu} q_{\sigma}-k^2 \delta_{\mu,\sigma}}{q^2-k^2-i0} F_{\mu}^{\sigma,ji}({\bf q})
 \,,
\label{Xi2}
\\
&&F_{\mu}^{\sigma,ji}({\bf q}) = \langle X, 1 {\cal F}_z,j | j_{\mu}(-{\bf q})|0 \rangle \langle 0|j^{\sigma}({\bf q}) |X, 1 {\cal F}_z,i \rangle\,. \nonumber
\end{eqnarray}
\end{subequations}
Note that the matrix $\Xi^{ji}$ is symmetric but not hermitian. Considering Eq.~(\ref{lambda2}) as an inhomogeneous system of linear equations on
\begin{equation}
\frac{\Lambda_i}{\omega_i-\omega-i0} \,,
\end{equation}
we can formally resolve it using Cramer's rule. This gives the following equation for the resonant frequencies:
\begin{equation}
\det{\left| \left| (\omega_j-\omega) \delta_{ji} + \Xi^{ji} \right| \right|}=0 \,.
\label{bigdet}
\end{equation}

Because of the symmetry all the matrix elements $\Xi^{ji}= \Xi$ are the same. The same refers to the unperturbed resonant frequencies: $\omega_i=\omega_j =\omega_0$. Therefore, in the left-hand side of Eq.~(\ref{bigdet}) we have a determinant of the matrix
\begin{equation}
 M_{ji} = (\omega_0-\omega) \delta_{ji} + \Xi \:.
\end{equation}
This allows one to rewrite Eq.~(\ref{bigdet}) as
\begin{equation}
(\omega_0-\omega+N \, \Xi) \,(\omega_0-\omega)^{N-1}=0 \,.
\end{equation}
In other words, out of $N=4$ valley-degenerate states excited by the light of given polarization 
only one gets frequency and radiative damping renormalizations, corresponding to the real and imaginary parts of $N \Xi$, respectively. They are both $N$ times as large as their single-valley counterparts. All the remaining $N-1$ states have no radiative decay and
become subradiant. For the decay rate of the ultrabright state in a PbX QD ($N=4$) we obtain
\begin{equation}
\frac{1}{\tau_r}
=\frac{32 \, e^2 \,k^3 }{27 \, \varepsilon_b \hbar} R^2 I_r^2 \,.
\label{result1}
\end{equation}
The ratio $e^2 k^3 / \varepsilon_b \hbar$ can be conveniently replaced by $\alpha \omega^3 \sqrt{\varepsilon_b}/c^2$, where $\alpha$ is the fine structure constant $e^2/c \hbar$.
In Appendix~\ref{sec:app_a} we show that the result $\tau^{-1}_r = N \tau^{-1}_0$ holds with allowance for the valley anisotropy.

Now the question arises: is the ultrabright state in a QD of multi-valley semiconductor superradiant? The pairwise interactions between emitters are known to destroy superradiance~\cite{Gross1982}, unless the system possesses additional symmetry leading to equivalence of the interaction energy 
for all the emitters. For example, in a Gedanken experiment proposed in Ref.~\citenum{Gross1982} this was achieved by arranging emitters to form a ring.
At first glance, if interactions are allowed between the valleys considered as emitters, then the interaction energy should be equal for all the valleys. However, if the shape of the QD has symmetry not lower than the symmetry of the crystal lattice, then the latter dictates that exciton states from different valleys form combinations representing basis functions of irreducible representations of the symmetry group. In Ref.~\citenum{Avdeev2020} the mechanism ensuring obedience to the lattice symmetry was called ``valley mixing''. 
While the Maxwell field tries to arrange emitters (valleys in our case) in a fully symmetric combinations, valley mixing favors combinations prescribed by the lattice symmetry. Furthermore, while the only ``conventional'' superradiant state accessible in linear regime is the lowest radiative state of $N$ emitters and electro-magnetic field promoting this state affects only its radiative decay rate, the ultrabright state is boosted by the longitudinal component of the Maxwell field and has energy higher than the energy of degenerate subradiant states as well as that of dark states of the direct and indirect excitons. Valley mixing distorts the ultrabright state and leads to brightening of subradiant and some dark states (originating from indirect excitons) via their admixture, while the state inheriting most properties 
of the ultrabright state is the highest state of the exciton multiplet~\cite{Avdeev2020}. 

\section{Calculation of the radiative lifetime for the ultrabright state}
\begin{figure}
  \centering{\includegraphics[width=0.9\linewidth]{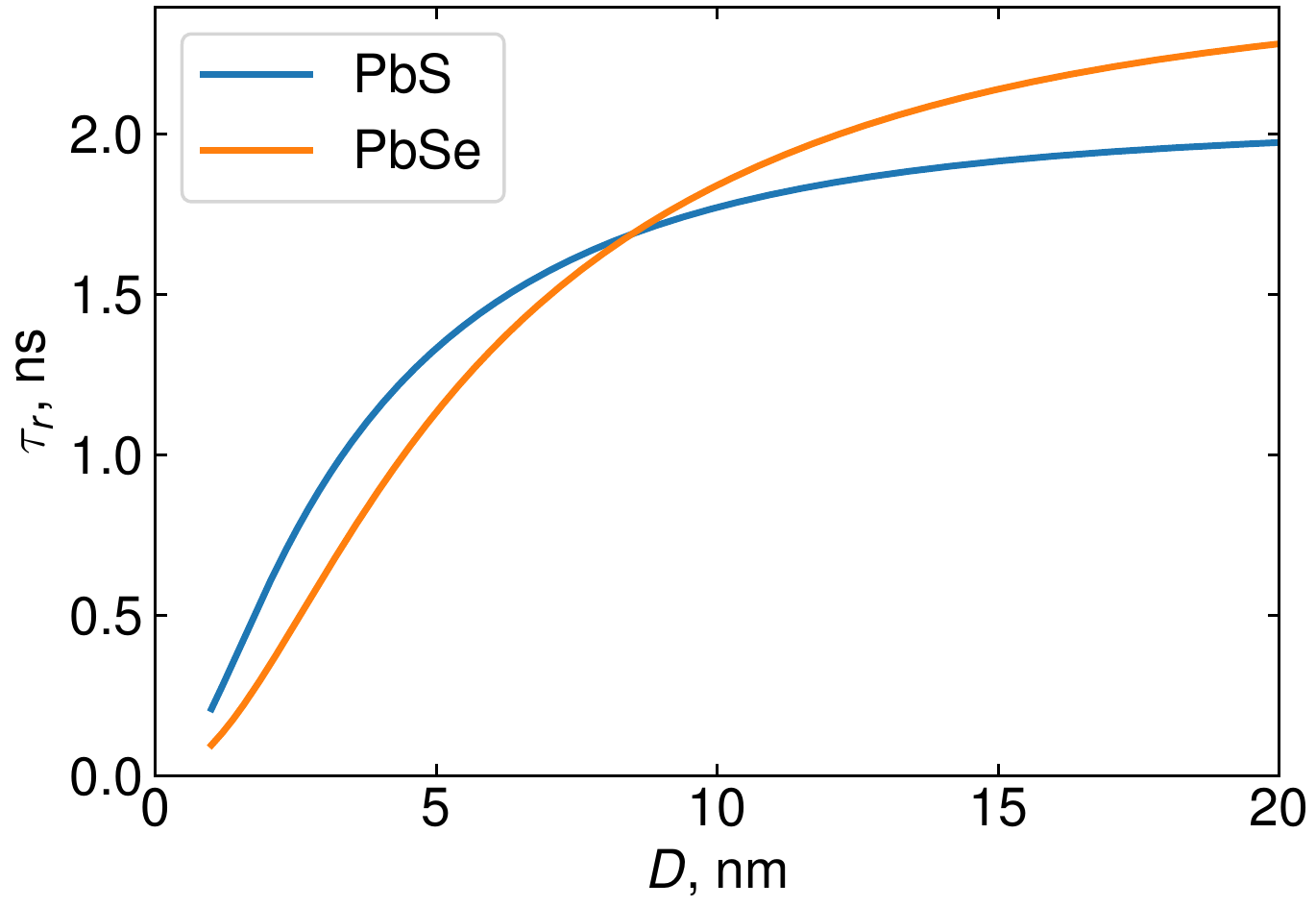}}
  \caption{
    Size dependence of the radiative decay time $\tau_r$, Eq.~\eqref{result1}, for the ultrabright state in PbS (blue line) and PbSe (orange line) QDs. For the calculations we use the parameters from Ref.~\citenum{Kang1997}. For PbS they are: $E_g=0.41$~eV, $\hbar v_0=3.09$~eV$\cdot$\AA, $\alpha_v=11.4$~eV$\cdot$\AA$^2$, $\alpha_c=9.52$~eV$\cdot$\AA$^2$, $\varepsilon_b=17$; for PbSe: $E_g=0.28$~eV, $\hbar v_0=3.15$~eV$\cdot$\AA, $\alpha_v=26.3$~eV$\cdot$\AA$^2$, $\alpha_c=14.9$~eV$\cdot$\AA$^2$, $\varepsilon_b=23$.
  }\label{fig}
\end{figure}

The size dependences of the radiative lifetimes for the ultrabright states in PbS and PbSe QDs obtained according to Eq.~(\ref{result1}) and using material parameters of Ref.\cite{Kang1997}  are shown in Fig.~\ref{fig}. 
They are in the nanosecond range.
These lifetimes should be compared to the experimentally observed photoluminescence decay times~\cite{Warner2004,Oron2009,Kigel2009} which are in the microsecond range. Several factors should be taken into account while performing such comparison. 

First, if the difference in background dielectric permittivities of the QD and its environment is taken into account then the radiative decay rate should be multiplied by the factor~\citep{goupalov2003}
\[
\frac{9 \, \varepsilon^2_{\rm out}}{(\varepsilon_{\rm b}+ 2 \, \varepsilon_{\rm out})^2} \,,
\]
where $\varepsilon_{\rm out}$ is the dielectric constant of the environment,
which reduces the radiative decay rate due to the large dielectric constants of PbX. 
The resulting radiative decay rate of the ultrabright state takes the form
\begin{equation}
\frac{1}{\tau_r}
=\frac{32 \sqrt{\varepsilon_{\rm b}} \alpha \omega^3}{3 c^2} \,
\frac{\varepsilon_{\mathrm{\rm out}}^2}{(\varepsilon_{\rm b}+ 2 \, \varepsilon_{\mathrm{\rm out}})^2} \, 
R^2 I_r^2 
\,,
\label{result1_dc}
\end{equation}
This dielectric contrast also leads to an additional contribution to the exchange splitting \cite{Avdeev2020}
\begin{equation}\label{eq:Xi_dc}
  \hbar \delta \omega_{\mathrm{dc}} = \frac49 \frac{e^2}{R} \frac{\varepsilon_{\mathrm b}-\varepsilon_{\mathrm{\rm out}}}{\varepsilon_{\mathrm b}\left(\varepsilon_{\mathrm b}+2\varepsilon_{\mathrm{\rm out}} \right) }
  I_r^2\,.
\end{equation}
This contribution for PbX QDs surrounded by a low dielectric constant medium is only few times smaller than the main contribution [cf. Eqs.~\eqref{Xideltaomega},~\eqref{Xi1}] and cannot be neglected. For PbS QDs embedded in the BK7 optical glass ($\varepsilon_{\mathrm{out}}=2.3$) the radiative lifetime is multiplied by a factor of $\sim$9.8 and the exchange splitting gets an about 20\% increase. 
For the colloidal QDs in hexane ($\varepsilon_{\mathrm{\rm out}}=1.94$) or in the air one could expect similar or larger factors.
Note that both Eqs.~(\ref{result1_dc}) and~(\ref{eq:Xi_dc}) can be derived within the formalism explained in Sec.~\ref{formalism}.

Second, the experimental photoluminescence dynamics can be influenced by more than one emitting state and involve nominally dark states with long lifetimes~\cite{Oron2009,Biadala2009}.

Finally, as mentioned at the end of Sec.~\ref{ultrabright},
the exciton states are greatly influenced by the valley mixing which leads to a redistribution of the oscillator strength of the ultrabright state 
among several available optically active states allowed by the symmetry~\cite{Avdeev2020}. In the next section this issue will be addressed in more details.

\section{Role of valley mixing}

In the isotropic model we considered thus far, the electron and hole ground states in a PbX QD, described in Sec.~\ref{ehground}, were eight-fold  spin- and valley-degenerate. The cubic symmetry of the crystal lattice dictates that the corresponding electron and hole energy levels should split into sublevels corresponding to irreducible representations of the point symmetry group of the QD. This symmetry group can coincide with the point group of the underlying rock-salt crystal lattice $O_h$ or have a lower symmetry. We will restrict our consideration by the QDs which have the same rotational symmetry as the crystal lattice but do not possess a center of inversion~\cite{Poddubny12}. Then the symmetry group of the QD is $T_d$ and both the electron and the hole ground levels split into two doublets corresponding to the irreducible representations $\Gamma_6$ and $\Gamma_7$ (below we use notation from Ref.~\citenum{Koster}) and one quadruplet corresponding to the irreducible representation $\Gamma_8$~\cite{Delerue2004b,Poddubny12,Avdeev2020}. As the resulting wave functions contain contributions from different valleys, we refer to these splittings as being caused by valley mixing. 

\begin{figure}[t]
\includegraphics[width=0.9\linewidth]{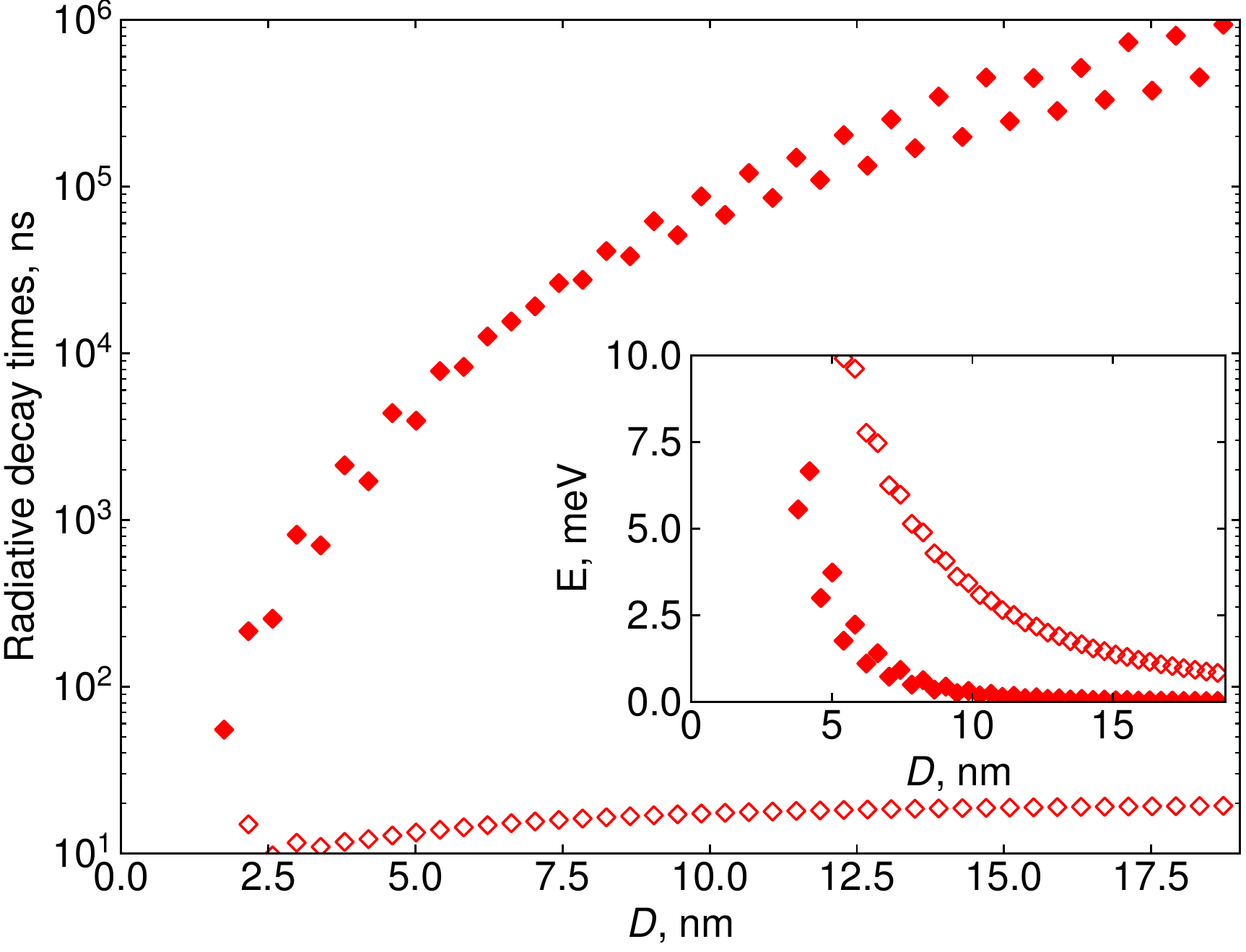}
\caption{
  Radiative lifetimes calculated taking into account both the valley mixing and the dielectric contrast for the lowest energy (solid diamonds) and highest energy (open diamonds) optically active exciton states as functions of the QD effective diameter for octahedron-shaped PbS QDs embedded in BK7 optical glass.
  Inset: energy splitting between the lowest energy (solid diamonds) and highest energy (open diamonds) optically active exciton states and the lowest dark  exciton state as a funciton of the QD effective diameter for octahedron-shaped PbS QDs embedded in BK7 optical glass.
}
\label{fig:VM_n_DC}
\end{figure}

When it comes to the exciton states, they cannot any longer be separated into the states of direct and indirect excitons. Instead, the exciton states can also be classified with respect to the irreducible representations of the $T_d$ group~\cite{Hu19,Avdeev2020}. It turns out that the 64-fold degenerate ground exciton level splits into 27 sublevels, 8 of which are optically active triplets corresponding to the irreducible representations $\Gamma_5$ of the group $T_d$~\cite{Hu19,Avdeev2020}. Thus, the oscillator strength of the ultrabright triplet of the isotropic model gets redistributed among all the eight
triplets of $\Gamma_5$ symmetry. 

Technically, one can include the eight states optically active in the ${\cal F}_z$ polarization with the energies not affected by the long-range electron-hole exchange interaction into Eq.~(\ref{t2}) and modify the derivation accordingly. 
Note that one can distinguish two sets of basis states. The first one is the basis of independent valleys which has been used to calculate the matrix elements~(\ref{Xi2}). The second one is the basis of direct products of irreducible representations of the group $T_d$, accounting for transformations of the single-particle wave functions. The latter basis diagonalizes the Hamiltoninan of non-interacting electron-hole pairs. For practical calculations it is convenient to transform this Hamiltonian to the basis of independent valleys. The explicit form of the transformation matrix is given in Ref.~\citenum{Avdeev2020}. 

In Fig.~\ref{fig:VM_n_DC} we show the calculated radiative lifetimes for the highest- and lowest-energy sublevels out of the eight optically active triplets as functions of the QD effective diameter. For small QDs, the lifetime of the lowest-energy optically active state, contributing to the low-temperature photoluminescence, is comparable with the lifetime of the highest-energy state inheriting most properties of the ultrabright state.  With the increase of QD size (and decrease of valley mixing) this lifetime rapidly grows and reaches the $\mu$s range. Calculations were performed in the framework of the extended \KP~model, where the splittings induced by the valley mixing were taken from the tight-binding results~\cite{Avdeev2020}.
For this particular calculation the QDs were chosen in the shape of octahedron.
Similar calculations of radiative lifetimes for QDs of various shapes show that, while the behavior of the lowest optical transition time as a functions of the QD size is smooth for QDs of the same shape, this time significantly
varies when going from one QD shape to another. In particular, for a 10~nm QD it is of the order of 0.3~$\mu$s for a cuboctahedral QD, 2~$\mu$s for a cubic QD and 50~$\mu$s for an octahedral QD, see Appendix~\ref{sec:app_b}.

\section{Conclusions}
When addressing optical transitions in nanostructures made of narrow-gap materials with strong interband coupling described by the non-diagonal part of the Dirac-like Hamiltonian, Eq.~(\ref{dimmock}), one is tempted to expect that the optical matrix element is proportional to the Fermi velocity $v_0$ of the gapless limit. We have demonstrated that this expectation is not right and one has to take into account the contribution of the massive terms to the interband matrix elements of the velocity operator. 

Taking into account multiple valleys leads to emergence of valley coherence when different valleys act as independent emitters in the reciprocal space and their symmetric combination becomes superradiant. The resulting ultrabright state of a PbX QD has a reduced radiative lifetime. However, cubic symmetry of the crystal lattice promotes different combinations of the valley states. This valley mixing leads to redistribution of the oscillator strength of the ultrabright state among 8 radiative triplets allowed by the symmetry. The radiative triplet having lowest energy is responsible for low-temperature photoluminescence and has radiative lifetime in the microsecond range in agreement with experimental findings.

\section*{Acknowledgments}
Authors thank M.M. Glazov for helpful discussions.
The work of SVG was supported by NSF through DMR-2100248.
The work of MON and ELI was funded by RFBR and CNRS according to the research project \#20-52-16303.
MON also thanks the Foundation for Advancement of Theoretical Physics and Mathematics ``BASIS''.

\begin{appendix}

\section{Superradiance in the \texorpdfstring{${\bf k}$}{k}-space. The case of anisotropic valleys}\label{sec:app_a}
Here we show that the superradiant regime is retained when the valley anisotropy is taken into account. 
Following Secs.~\ref{formalism} and~\ref{ultrabright}, we consider a state of the exciton confined in a QD originating from the valley oriented along $[111]$. Let us use the Cartesian coordinate system with 
\begin{equation}\label{eq:cs_v}
  x_1 || [11\bar2]\,,\;\;\;
  y_1 || [\bar110]\,,\;\;\;
  z_1 || [111]\,.
\end{equation}
The exciton ground state is formed by the optically inactive sublevel $\ket{exc,0}$, two sublevels of the dublet $\ket{X, 1 {\cal F}_{x_1}}$, $\ket{X, 1 {\cal F}_{y_1}}$ optically active in polarizations ${\bf e} || x_1$ and ${\bf e} || y_1$, and the sublevel $\ket{X, 1 {\cal F}_{z_1}}$ active in polarization ${\bf e} || z_1$.

Due to the axial symmetry of a single valley, the matrix elements for the photoexcitation of the active sublevels may be written as
\begin{equation}
\begin{split}
  M_{X, 1 {\cal F}_{x_1},i=1} = v_{\perp} e_{x_1}\,,\\
  M_{X, 1 {\cal F}_{y_1},i=1} = v_{\perp} e_{y_1}\,,\\
  M_{X, 1 {\cal F}_{z_1},i=1} = v_{||} e_{z_1}\,,
\end{split}
\end{equation}
or, in the crystallographic system of coordinates $x|| [100]$, $y|| [010]$, $z || [001]$, 
\begin{equation}
\begin{split}
  M_{X, 1 {\cal F}_{x1},i=1} &= v_{\perp} \frac1{\sqrt6} \left( e_{x} + e_{y} -2 e_{z}\right)\,,\\
  M_{X, 1 {\cal F}_{y1},i=1} &= v_{\perp} \frac1{\sqrt2} \left(-e_{x} + e_{y}         \right)\,,\\
  M_{X, 1 {\cal F}_{z1},i=1} &= v_{||} \frac1{\sqrt3} \left( e_{x} + e_{y} + e_{z}\right)\,,
\end{split}
\end{equation}
where $v_{\perp}$, $v_{||}$ are the optical matrix elements.

Linear combination of the three excitonic states $\ket{X, 1 {\cal F}_{x_1}}$, $\ket{X, 1 {\cal F}_{y_1}}$, $\ket{X, 1 {\cal F}_{z_1}}$ polarized along $z$ axis has the form
\begin{equation}
\begin{split}
\ket{X, 1 {\cal F}_z} &= C \left( \frac{\sqrt3}{v_{||}} \ket{X, 1 {\cal F}_{z_1}} - \frac{\sqrt6}{v_{\perp}} \ket{X, 1 {\cal F}_{x_1}} \right)\,,\\
 C &= \frac1{\sqrt3}\frac{v_{||}v_{\perp}}{\sqrt{v_{\perp}^2+2v_{||}^2}}\,,
\end{split}
\end{equation}
where we used the relation between the valley coordinate system \eqref{eq:cs_v} and the crystallographic coordinate system.
One may check that 
\begin{multline}\label{eq:M1z}
M_{X, 1 {\cal F}_z,i=1} = C\left[
       \frac{\sqrt3}{v_{||}} v_{||} \frac1{\sqrt3} (e_x+e_y+e_z)
       \right.\\\left.
      -\frac{\sqrt6}{v_{\perp}} v_{\perp} \frac1{\sqrt6} (e_x+e_y-2e_z)
       \right] = 3C e_z\,.
\end{multline}

Exciton states $\ket{X, 1 {\cal F}_z,i}$ in the three other valleys $i=2,3,4$ are obtained by applying the operations of $C_4$, $C_4^2$, $C_4^3$ to the state $\ket{X, 1 {\cal F}_z,i=1}$. Since $z$ is invariant under these three operations, from \eqref{eq:M1z} one may obtain that, for all four 
excitons $\ket{X, 1 {\cal F}_z,i}$ ($i=1,2,3,4$), the matrix elements are equal: 
\begin{equation}\label{eq:Miz}
  M_{X, 1 {\cal F}_z,i} = 3C e_z\,.
\end{equation}

It follows from the above considerations that the combination optically active in polarization ${\bf e}|| z$ is 
\begin{multline}
  \ket{X, 1 {\cal F}_z,\Gamma_1} = \frac12 \left( \ket{X, 1 {\cal F}_z,1} + \ket{X, 1 {\cal F}_z,2} \right.
  \\+ \left. \ket{X, 1 {\cal F}_z,3} + \ket{X, 1 {\cal F}_z,4} \right)\,,
\end{multline}
\[  
  M_{X, 1 {\cal F}_z,\Gamma_1} = 6 C e_z\,.
\]
The three states 
\begin{eqnarray}
  \ket{X, 1 {\cal F}_z,\Gamma_5^x} &=& \frac12 \left(  \ket{X, 1 {\cal F}_z,1} - \ket{X, 1 {\cal F}_z,2} \right.\\
  && \left. - \ket{X, 1 {\cal F}_z,3} + \ket{X, 1 {\cal F}_z,4} \right)\,,\nonumber\\
  \ket{X, 1 {\cal F}_z,\Gamma_5^y} &=& \frac12 \left(  \ket{X, 1 {\cal F}_z,1} + \ket{X, 1 {\cal F}_z,2} \right. \\
  && \left. - \ket{X, 1 {\cal F}_z,3} - \ket{X, 1 {\cal F}_z,4} \right)\,,\nonumber\\
  \ket{X, 1 {\cal F}_z,\Gamma_5^z} &=& \frac12 \left(  \ket{X, 1 {\cal F}_z,1} - \ket{X, 1 {\cal F}_z,2} \right. \\
  && \left. + \ket{X, 1 {\cal F}_z,3} - \ket{X, 1 {\cal F}_z,4} \right)\,\nonumber
\end{eqnarray}
are optically inactive.

The action of the operations $C_3$ and $C_3^2$ on the state $\ket{X, 1 {\cal F}_z,\Gamma_1}$ gives the states $\ket{X, 1 {\cal F}_x,\Gamma_1}$ and $\ket{X, 1 {\cal F}_y,\Gamma_1}$, polarized along $x$ and $y$, respectively.

To conclude, out of the 16 states of the direct exciton only three are optically active: $\ket{X, 1 {\cal F}_x,\Gamma_1}$, $\ket{X, 1 {\cal F}_y,\Gamma_1}$, and $\ket{X, 1 {\cal F}_z,\Gamma_1}$. These states transform according to the $\Gamma_5 \otimes \Gamma_1=\Gamma_5$ representation of the group $T_d$.

\section{Radiative times for QDs of different shape.}\label{sec:app_b}

\begin{figure*}[htb]
\includegraphics[width=0.8\textwidth]{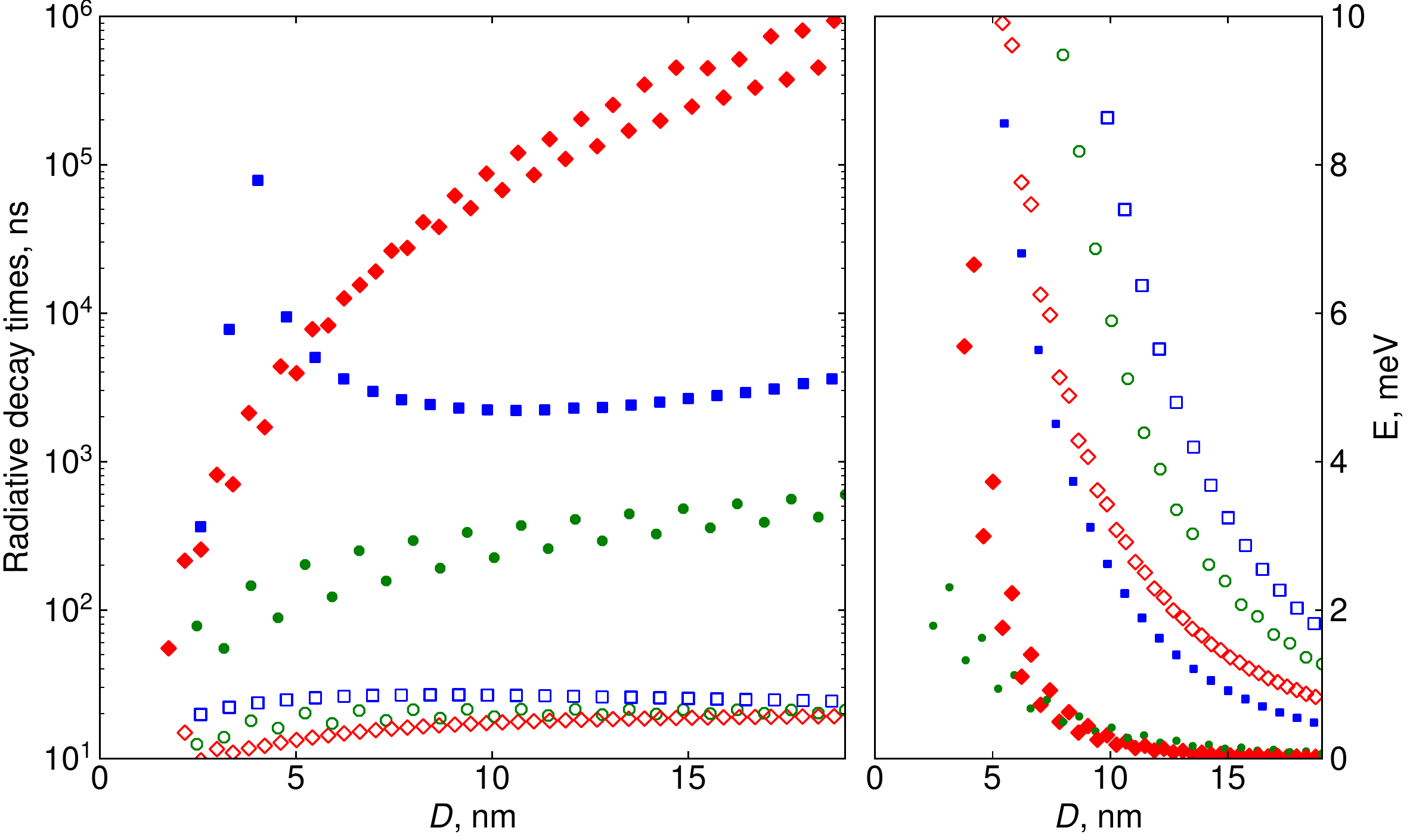}
\caption{
  Left panel shows the radiative lifetimes calculated taking into account both the valley mixing and the dielectric contrast for the lowest energy (solid symbols) and highest energy (open symbols) optically active exciton states as functions of the QD effective diameter for PbS QDs of different shape embedded in BK7 optical glass.
  Right panel shows the energy splitting between the lowest energy (solid symbols) and highest energy (open symbols) optically active exciton states and the lowest dark exciton state as a funciton of the QD effective diameter for PbS QDs embedded in BK7 optical glass.
  Different symbol shapes and colors encode the QD shape: red diamonds, green octagons and blue squares show the data for octaherdal, cuboctahedral and cubic QDs, respectively.
}
\label{fig:VM_n_DC_all}
\end{figure*}

In Fig.~\ref{fig:VM_n_DC_all} we show the calculated radiative lifetimes for the highest- and lowest-energy sublevels out of the eight optically active triplets as functions of the QD effective diameter. Calculations were performed in the framework of the extended \KP~model, where the splittings induced by the valley mixing were taken from the tight-binding results~\cite{Avdeev2020}. Here we present the calculations of radiative lifetimes for QDs of various shapes.

\end{appendix}

\bibliography{PbS}

\begin{thebibliography}{30}%
\makeatletter
\providecommand \@ifxundefined [1]{%
 \@ifx{#1\undefined}
}%
\providecommand \@ifnum [1]{%
 \ifnum #1\expandafter \@firstoftwo
 \else \expandafter \@secondoftwo
 \fi
}%
\providecommand \@ifx [1]{%
 \ifx #1\expandafter \@firstoftwo
 \else \expandafter \@secondoftwo
 \fi
}%
\providecommand \natexlab [1]{#1}%
\providecommand \enquote  [1]{``#1''}%
\providecommand \bibnamefont  [1]{#1}%
\providecommand \bibfnamefont [1]{#1}%
\providecommand \citenamefont [1]{#1}%
\providecommand \href@noop [0]{\@secondoftwo}%
\providecommand \href [0]{\begingroup \@sanitize@url \@href}%
\providecommand \@href[1]{\@@startlink{#1}\@@href}%
\providecommand \@@href[1]{\endgroup#1\@@endlink}%
\providecommand \@sanitize@url [0]{\catcode `\\12\catcode `\$12\catcode
  `\&12\catcode `\#12\catcode `\^12\catcode `\_12\catcode `\%12\relax}%
\providecommand \@@startlink[1]{}%
\providecommand \@@endlink[0]{}%
\providecommand \url  [0]{\begingroup\@sanitize@url \@url }%
\providecommand \@url [1]{\endgroup\@href {#1}{\urlprefix }}%
\providecommand \urlprefix  [0]{URL }%
\providecommand \Eprint [0]{\href }%
\providecommand \doibase [0]{https://doi.org/}%
\providecommand \selectlanguage [0]{\@gobble}%
\providecommand \bibinfo  [0]{\@secondoftwo}%
\providecommand \bibfield  [0]{\@secondoftwo}%
\providecommand \translation [1]{[#1]}%
\providecommand \BibitemOpen [0]{}%
\providecommand \bibitemStop [0]{}%
\providecommand \bibitemNoStop [0]{.\EOS\space}%
\providecommand \EOS [0]{\spacefactor3000\relax}%
\providecommand \BibitemShut  [1]{\csname bibitem#1\endcsname}%
\let\auto@bib@innerbib\@empty
\bibitem [{\citenamefont {Sun}\ \emph {et~al.}(2012)\citenamefont {Sun},
  \citenamefont {Choi}, \citenamefont {Stachnik}, \citenamefont {Bartnik},
  \citenamefont {Hyun}, \citenamefont {Malliaras}, \citenamefont {Hanrath},\
  and\ \citenamefont {Wise}}]{Sun2012}%
  \BibitemOpen
  \bibfield  {author} {\bibinfo {author} {\bibfnamefont {L.}~\bibnamefont
  {Sun}}, \bibinfo {author} {\bibfnamefont {J.~J.}\ \bibnamefont {Choi}},
  \bibinfo {author} {\bibfnamefont {D.}~\bibnamefont {Stachnik}}, \bibinfo
  {author} {\bibfnamefont {A.~C.}\ \bibnamefont {Bartnik}}, \bibinfo {author}
  {\bibfnamefont {B.-R.}\ \bibnamefont {Hyun}}, \bibinfo {author}
  {\bibfnamefont {G.~G.}\ \bibnamefont {Malliaras}}, \bibinfo {author}
  {\bibfnamefont {T.}~\bibnamefont {Hanrath}},\ and\ \bibinfo {author}
  {\bibfnamefont {F.~W.}\ \bibnamefont {Wise}},\ }\bibfield  {title} {\bibinfo
  {title} {Bright infrared quantum-dot light-emitting diodes through inter-dot
  spacing control},\ }\href {https://doi.org/10.1038/nnano.2012.63} {\bibfield
  {journal} {\bibinfo  {journal} {Nature Nanotech.}\ }\textbf {\bibinfo
  {volume} {7}},\ \bibinfo {pages} {369} (\bibinfo {year} {2012})}\BibitemShut
  {NoStop}%
\bibitem [{\citenamefont {Sukhovatkin}\ \emph {et~al.}(2009)\citenamefont
  {Sukhovatkin}, \citenamefont {Hinds}, \citenamefont {Brzozowski},\ and\
  \citenamefont {Sargent}}]{Sukhovatkin09}%
  \BibitemOpen
  \bibfield  {author} {\bibinfo {author} {\bibfnamefont {V.}~\bibnamefont
  {Sukhovatkin}}, \bibinfo {author} {\bibfnamefont {S.}~\bibnamefont {Hinds}},
  \bibinfo {author} {\bibfnamefont {L.}~\bibnamefont {Brzozowski}},\ and\
  \bibinfo {author} {\bibfnamefont {E.~H.}\ \bibnamefont {Sargent}},\
  }\bibfield  {title} {\bibinfo {title} {Colloidal quantum-dot photodetectors
  exploiting multiexciton generation},\ }\href
  {https://doi.org/10.1126/science.1173812} {\bibfield  {journal} {\bibinfo
  {journal} {Science}\ }\textbf {\bibinfo {volume} {324}},\ \bibinfo {pages}
  {1542} (\bibinfo {year} {2009})}\BibitemShut {NoStop}%
\bibitem [{\citenamefont {Tisdale}\ \emph {et~al.}(2010)\citenamefont
  {Tisdale}, \citenamefont {Williams}, \citenamefont {Timp}, \citenamefont
  {Norris}, \citenamefont {Aydil},\ and\ \citenamefont {Zhu}}]{Tisdale2010}%
  \BibitemOpen
  \bibfield  {author} {\bibinfo {author} {\bibfnamefont {W.~A.}\ \bibnamefont
  {Tisdale}}, \bibinfo {author} {\bibfnamefont {K.~J.}\ \bibnamefont
  {Williams}}, \bibinfo {author} {\bibfnamefont {B.~A.}\ \bibnamefont {Timp}},
  \bibinfo {author} {\bibfnamefont {D.~J.}\ \bibnamefont {Norris}}, \bibinfo
  {author} {\bibfnamefont {E.~S.}\ \bibnamefont {Aydil}},\ and\ \bibinfo
  {author} {\bibfnamefont {X.-Y.}\ \bibnamefont {Zhu}},\ }\bibfield  {title}
  {\bibinfo {title} {Hot-electron transfer from semiconductor nanocrystals},\
  }\href {https://doi.org/10.1126/science.1185509} {\bibfield  {journal}
  {\bibinfo  {journal} {Science}\ }\textbf {\bibinfo {volume} {328}},\ \bibinfo
  {pages} {1543} (\bibinfo {year} {2010})}\BibitemShut {NoStop}%
\bibitem [{\citenamefont {Gao}\ \emph {et~al.}(2020)\citenamefont {Gao},
  \citenamefont {Quan}, \citenamefont {de~Arquer}, \citenamefont {Zhao},
  \citenamefont {Munir}, \citenamefont {Proppe}, \citenamefont
  {Quintero-Bermudez}, \citenamefont {Zou}, \citenamefont {Yang}, \citenamefont
  {Saidaminov}, \citenamefont {Voznyy}, \citenamefont {Kinge}, \citenamefont
  {Lu}, \citenamefont {Kelley}, \citenamefont {Amassian}, \citenamefont
  {Tang},\ and\ \citenamefont {Sargent}}]{Gao2020}%
  \BibitemOpen
  \bibfield  {author} {\bibinfo {author} {\bibfnamefont {L.}~\bibnamefont
  {Gao}}, \bibinfo {author} {\bibfnamefont {L.~N.}\ \bibnamefont {Quan}},
  \bibinfo {author} {\bibfnamefont {F.~P.~G.}\ \bibnamefont {de~Arquer}},
  \bibinfo {author} {\bibfnamefont {Y.}~\bibnamefont {Zhao}}, \bibinfo {author}
  {\bibfnamefont {R.}~\bibnamefont {Munir}}, \bibinfo {author} {\bibfnamefont
  {A.}~\bibnamefont {Proppe}}, \bibinfo {author} {\bibfnamefont
  {R.}~\bibnamefont {Quintero-Bermudez}}, \bibinfo {author} {\bibfnamefont
  {C.}~\bibnamefont {Zou}}, \bibinfo {author} {\bibfnamefont {Z.}~\bibnamefont
  {Yang}}, \bibinfo {author} {\bibfnamefont {M.~I.}\ \bibnamefont
  {Saidaminov}}, \bibinfo {author} {\bibfnamefont {O.}~\bibnamefont {Voznyy}},
  \bibinfo {author} {\bibfnamefont {S.}~\bibnamefont {Kinge}}, \bibinfo
  {author} {\bibfnamefont {Z.}~\bibnamefont {Lu}}, \bibinfo {author}
  {\bibfnamefont {S.~O.}\ \bibnamefont {Kelley}}, \bibinfo {author}
  {\bibfnamefont {A.}~\bibnamefont {Amassian}}, \bibinfo {author}
  {\bibfnamefont {J.}~\bibnamefont {Tang}},\ and\ \bibinfo {author}
  {\bibfnamefont {E.~H.}\ \bibnamefont {Sargent}},\ }\bibfield  {title}
  {\bibinfo {title} {Efficient near-infrared light-emitting diodes based on
  quantum dots in layered perovskite},\ }\href
  {https://doi.org/10.1038/s41566-019-0577-1} {\bibfield  {journal} {\bibinfo
  {journal} {Nature Photonics}\ }\textbf {\bibinfo {volume} {14}},\ \bibinfo
  {pages} {227} (\bibinfo {year} {2020})}\BibitemShut {NoStop}%
\bibitem [{\citenamefont {Lu}\ \emph {et~al.}(2020)\citenamefont {Lu},
  \citenamefont {Huang}, \citenamefont {Martinez}, \citenamefont {Johnson},
  \citenamefont {Luther},\ and\ \citenamefont {Beard}}]{Lu2020}%
  \BibitemOpen
  \bibfield  {author} {\bibinfo {author} {\bibfnamefont {H.}~\bibnamefont
  {Lu}}, \bibinfo {author} {\bibfnamefont {Z.}~\bibnamefont {Huang}}, \bibinfo
  {author} {\bibfnamefont {M.~S.}\ \bibnamefont {Martinez}}, \bibinfo {author}
  {\bibfnamefont {J.~C.}\ \bibnamefont {Johnson}}, \bibinfo {author}
  {\bibfnamefont {J.~M.}\ \bibnamefont {Luther}},\ and\ \bibinfo {author}
  {\bibfnamefont {M.~C.}\ \bibnamefont {Beard}},\ }\bibfield  {title} {\bibinfo
  {title} {Transforming energy using quantum dots},\ }\href
  {https://doi.org/10.1039/C9EE03930A} {\bibfield  {journal} {\bibinfo
  {journal} {Energy Environ. Sci.}\ }\textbf {\bibinfo {volume} {13}},\
  \bibinfo {pages} {1347} (\bibinfo {year} {2020})}\BibitemShut {NoStop}%
\bibitem [{\citenamefont {Kong}\ \emph {et~al.}(2016)\citenamefont {Kong},
  \citenamefont {Chen}, \citenamefont {Fang}, \citenamefont {Heath},
  \citenamefont {Wo}, \citenamefont {Wang}, \citenamefont {Li}, \citenamefont
  {Guo}, \citenamefont {Evans}, \citenamefont {Chen},\ and\ \citenamefont
  {Zhou}}]{Kong2016}%
  \BibitemOpen
  \bibfield  {author} {\bibinfo {author} {\bibfnamefont {Y.}~\bibnamefont
  {Kong}}, \bibinfo {author} {\bibfnamefont {J.}~\bibnamefont {Chen}}, \bibinfo
  {author} {\bibfnamefont {H.}~\bibnamefont {Fang}}, \bibinfo {author}
  {\bibfnamefont {G.}~\bibnamefont {Heath}}, \bibinfo {author} {\bibfnamefont
  {Y.}~\bibnamefont {Wo}}, \bibinfo {author} {\bibfnamefont {W.}~\bibnamefont
  {Wang}}, \bibinfo {author} {\bibfnamefont {Y.}~\bibnamefont {Li}}, \bibinfo
  {author} {\bibfnamefont {Y.}~\bibnamefont {Guo}}, \bibinfo {author}
  {\bibfnamefont {S.~D.}\ \bibnamefont {Evans}}, \bibinfo {author}
  {\bibfnamefont {S.}~\bibnamefont {Chen}},\ and\ \bibinfo {author}
  {\bibfnamefont {D.}~\bibnamefont {Zhou}},\ }\bibfield  {title} {\bibinfo
  {title} {Highly fluorescent ribonuclease-a-encapsulated lead sulfide quantum
  dots for ultrasensitive fluorescence in vivo imaging in the second
  near-infrared window},\ }\href
  {https://doi.org/10.1021/acs.chemmater.6b00208} {\bibfield  {journal}
  {\bibinfo  {journal} {Chemistry of Materials}\ }\textbf {\bibinfo {volume}
  {28}},\ \bibinfo {pages} {3041} (\bibinfo {year} {2016})}\BibitemShut
  {NoStop}%
\bibitem [{\citenamefont {Zhang}\ \emph {et~al.}(2018)\citenamefont {Zhang},
  \citenamefont {Yue}, \citenamefont {Cui}, \citenamefont {Ma}, \citenamefont
  {Wan}, \citenamefont {Wang}, \citenamefont {Zhu}, \citenamefont {Zhou},
  \citenamefont {Kuang}, \citenamefont {Zhong}, \citenamefont {Pang},\ and\
  \citenamefont {Dai}}]{Zhang2018}%
  \BibitemOpen
  \bibfield  {author} {\bibinfo {author} {\bibfnamefont {M.}~\bibnamefont
  {Zhang}}, \bibinfo {author} {\bibfnamefont {J.}~\bibnamefont {Yue}}, \bibinfo
  {author} {\bibfnamefont {R.}~\bibnamefont {Cui}}, \bibinfo {author}
  {\bibfnamefont {Z.}~\bibnamefont {Ma}}, \bibinfo {author} {\bibfnamefont
  {H.}~\bibnamefont {Wan}}, \bibinfo {author} {\bibfnamefont {F.}~\bibnamefont
  {Wang}}, \bibinfo {author} {\bibfnamefont {S.}~\bibnamefont {Zhu}}, \bibinfo
  {author} {\bibfnamefont {Y.}~\bibnamefont {Zhou}}, \bibinfo {author}
  {\bibfnamefont {Y.}~\bibnamefont {Kuang}}, \bibinfo {author} {\bibfnamefont
  {Y.}~\bibnamefont {Zhong}}, \bibinfo {author} {\bibfnamefont {D.-W.}\
  \bibnamefont {Pang}},\ and\ \bibinfo {author} {\bibfnamefont
  {H.}~\bibnamefont {Dai}},\ }\bibfield  {title} {\bibinfo {title} {Bright
  quantum dots emitting at $\sim$1, 600 nm in the {NIR}-{IIb} window for deep
  tissue fluorescence imaging},\ }\href
  {https://doi.org/10.1073/pnas.1806153115} {\bibfield  {journal} {\bibinfo
  {journal} {Proceedings of the National Academy of Sciences}\ }\textbf
  {\bibinfo {volume} {115}},\ \bibinfo {pages} {6590} (\bibinfo {year}
  {2018})}\BibitemShut {NoStop}%
\bibitem [{\citenamefont {Xia}\ \emph {et~al.}(2021)\citenamefont {Xia},
  \citenamefont {Gevers}, \citenamefont {Fognini}, \citenamefont {Mok},
  \citenamefont {Li}, \citenamefont {Akbari}, \citenamefont {Zadeh},
  \citenamefont {Qin-Dregely},\ and\ \citenamefont {Xu}}]{Xia2021}%
  \BibitemOpen
  \bibfield  {author} {\bibinfo {author} {\bibfnamefont {F.}~\bibnamefont
  {Xia}}, \bibinfo {author} {\bibfnamefont {M.}~\bibnamefont {Gevers}},
  \bibinfo {author} {\bibfnamefont {A.}~\bibnamefont {Fognini}}, \bibinfo
  {author} {\bibfnamefont {A.~T.}\ \bibnamefont {Mok}}, \bibinfo {author}
  {\bibfnamefont {B.}~\bibnamefont {Li}}, \bibinfo {author} {\bibfnamefont
  {N.}~\bibnamefont {Akbari}}, \bibinfo {author} {\bibfnamefont {I.~E.}\
  \bibnamefont {Zadeh}}, \bibinfo {author} {\bibfnamefont {J.}~\bibnamefont
  {Qin-Dregely}},\ and\ \bibinfo {author} {\bibfnamefont {C.}~\bibnamefont
  {Xu}},\ }\bibfield  {title} {\bibinfo {title} {Short-wave infrared confocal
  fluorescence imaging of deep mouse brain with a superconducting nanowire
  single-photon detector},\ }\href
  {https://doi.org/10.1021/acsphotonics.1c01018} {\bibfield  {journal}
  {\bibinfo  {journal} {{ACS} Photonics}\ }\textbf {\bibinfo {volume} {8}},\
  \bibinfo {pages} {2800} (\bibinfo {year} {2021})}\BibitemShut {NoStop}%
\bibitem [{\citenamefont {Kang}\ and\ \citenamefont {Wise}(1997)}]{Kang1997}%
  \BibitemOpen
  \bibfield  {author} {\bibinfo {author} {\bibfnamefont {I.}~\bibnamefont
  {Kang}}\ and\ \bibinfo {author} {\bibfnamefont {F.~W.}\ \bibnamefont
  {Wise}},\ }\bibfield  {title} {\bibinfo {title} {Electronic structure and
  optical properties of {PbS} and {PbSe} quantum dots},\ }\href
  {https://doi.org/10.1364/JOSAB.14.001632} {\bibfield  {journal} {\bibinfo
  {journal} {J. Opt. Soc. Am. B}\ }\textbf {\bibinfo {volume} {14}},\ \bibinfo
  {pages} {1632} (\bibinfo {year} {1997})}\BibitemShut {NoStop}%
\bibitem [{\citenamefont {Berestetskii}\ \emph {et~al.}(1982)\citenamefont
  {Berestetskii}, \citenamefont {Lifshitz},\ and\ \citenamefont
  {Pitaevski{\u\i}}}]{Landau4_book}%
  \BibitemOpen
  \bibfield  {author} {\bibinfo {author} {\bibfnamefont {V.}~\bibnamefont
  {Berestetskii}}, \bibinfo {author} {\bibfnamefont {E.}~\bibnamefont
  {Lifshitz}},\ and\ \bibinfo {author} {\bibfnamefont {L.}~\bibnamefont
  {Pitaevski{\u\i}}},\ }\href@noop {} {\emph {\bibinfo {title} {Quantum
  Electrodynamics}}},\ Course of theoretical physics\ (\bibinfo  {publisher}
  {Butterworth-Heinemann},\ \bibinfo {year} {1982})\BibitemShut {NoStop}%
\bibitem [{\citenamefont {Lifshitz}\ and\ \citenamefont
  {Pitaevski\u{\i}}(1980)}]{Landau9_book}%
  \BibitemOpen
  \bibfield  {author} {\bibinfo {author} {\bibfnamefont {E.~M.}\ \bibnamefont
  {Lifshitz}}\ and\ \bibinfo {author} {\bibfnamefont {L.~P.}\ \bibnamefont
  {Pitaevski\u{\i}}},\ }\href@noop {} {\emph {\bibinfo {title} {Statistical
  Physics, Part 2: Theory of the Condensed State}}}\ (\bibinfo  {publisher}
  {Pergamon Press},\ \bibinfo {address} {Oxford},\ \bibinfo {year} {1980})\
  pp.\ \bibinfo {pages} {xi+387},\ \bibinfo {note} {course of Theoretical
  Physics , Vol. 9. Translated from the Russian by J. B. Sykes and M. J.
  Kearsley.}\BibitemShut {Stop}%
\bibitem [{\citenamefont {Blount}(1962)}]{Blount1962}%
  \BibitemOpen
  \bibfield  {author} {\bibinfo {author} {\bibfnamefont {E.~I.}\ \bibnamefont
  {Blount}},\ }\bibfield  {title} {\bibinfo {title} {Bloch electrons in a
  magnetic field},\ }\href {https://doi.org/10.1103/physrev.126.1636}
  {\bibfield  {journal} {\bibinfo  {journal} {Physical Review}\ }\textbf
  {\bibinfo {volume} {126}},\ \bibinfo {pages} {1636} (\bibinfo {year}
  {1962})}\BibitemShut {NoStop}%
\bibitem [{\citenamefont {Toyozawa}(2003)}]{toyozawa}%
  \BibitemOpen
  \bibfield  {author} {\bibinfo {author} {\bibfnamefont {Y.}~\bibnamefont
  {Toyozawa}},\ }\href@noop {} {\emph {\bibinfo {title} {Optical processes in
  solids}}}\ (\bibinfo  {publisher} {Cambridge University Press},\ \bibinfo
  {address} {Cambridge, UK New York},\ \bibinfo {year} {2003})\BibitemShut
  {NoStop}%
\bibitem [{\citenamefont {Cohen-Tannoudji}\ \emph {et~al.}(1989)\citenamefont
  {Cohen-Tannoudji}, \citenamefont {Dupont-Roc},\ and\ \citenamefont
  {Grynberg}}]{tannoudji1989}%
  \BibitemOpen
  \bibfield  {author} {\bibinfo {author} {\bibfnamefont {C.}~\bibnamefont
  {Cohen-Tannoudji}}, \bibinfo {author} {\bibfnamefont {J.}~\bibnamefont
  {Dupont-Roc}},\ and\ \bibinfo {author} {\bibfnamefont {G.}~\bibnamefont
  {Grynberg}},\ }\href@noop {} {\emph {\bibinfo {title} {Photons and atoms :
  introduction to quantum electrodynamics}}}\ (\bibinfo  {publisher} {Wiley},\
  \bibinfo {address} {New York},\ \bibinfo {year} {1989})\BibitemShut {NoStop}%
\bibitem [{\citenamefont {Avdeev}\ \emph {et~al.}(2020)\citenamefont {Avdeev},
  \citenamefont {Nestoklon},\ and\ \citenamefont {Goupalov}}]{Avdeev2020}%
  \BibitemOpen
  \bibfield  {author} {\bibinfo {author} {\bibfnamefont {I.~D.}\ \bibnamefont
  {Avdeev}}, \bibinfo {author} {\bibfnamefont {M.~O.}\ \bibnamefont
  {Nestoklon}},\ and\ \bibinfo {author} {\bibfnamefont {S.~V.}\ \bibnamefont
  {Goupalov}},\ }\bibfield  {title} {\bibinfo {title} {Exciton fine structure
  in lead chalcogenide quantum dots: Valley mixing and crucial role of
  intervalley electron{\textendash}hole exchange},\ }\href
  {https://doi.org/10.1021/acs.nanolett.0c03937} {\bibfield  {journal}
  {\bibinfo  {journal} {Nano Lett.}\ }\textbf {\bibinfo {volume} {20}},\
  \bibinfo {pages} {8897} (\bibinfo {year} {2020})}\BibitemShut {NoStop}%
\bibitem [{\citenamefont {Agranovich}\ and\ \citenamefont
  {Ginzburg}(1984)}]{Agranovich}%
  \BibitemOpen
  \bibfield  {author} {\bibinfo {author} {\bibfnamefont {V.~M.}\ \bibnamefont
  {Agranovich}}\ and\ \bibinfo {author} {\bibfnamefont {V.~L.}\ \bibnamefont
  {Ginzburg}},\ }\href@noop {} {\emph {\bibinfo {title} {Crystal Optics with
  Spatial Dispersion, and Excitons}}}\ (\bibinfo  {publisher}
  {Springer-Verlag},\ \bibinfo {year} {1984})\BibitemShut {NoStop}%
\bibitem [{\citenamefont {Gupalov}\ \emph {et~al.}(1998)\citenamefont
  {Gupalov}, \citenamefont {Ivchenko},\ and\ \citenamefont
  {Kavokin}}]{Gupalov1998}%
  \BibitemOpen
  \bibfield  {author} {\bibinfo {author} {\bibfnamefont {S.~V.}\ \bibnamefont
  {Gupalov}}, \bibinfo {author} {\bibfnamefont {E.~L.}\ \bibnamefont
  {Ivchenko}},\ and\ \bibinfo {author} {\bibfnamefont {A.~V.}\ \bibnamefont
  {Kavokin}},\ }\bibfield  {title} {\bibinfo {title} {Fine structure of
  localized exciton levels in quantum wells},\ }\href
  {https://doi.org/10.1134/1.558441} {\bibfield  {journal} {\bibinfo  {journal}
  {Journal of Experimental and Theoretical Physics}\ }\textbf {\bibinfo
  {volume} {86}},\ \bibinfo {pages} {388} (\bibinfo {year} {1998})}\BibitemShut
  {NoStop}%
\bibitem [{\citenamefont {Cho}(1999)}]{Cho1999}%
  \BibitemOpen
  \bibfield  {author} {\bibinfo {author} {\bibfnamefont {K.}~\bibnamefont
  {Cho}},\ }\bibfield  {title} {\bibinfo {title} {Mechanisms for {LT} splitting
  of polarization waves: a link between electron-hole exchange interaction and
  depolarization shift},\ }\href {https://doi.org/10.1143/jpsj.68.683}
  {\bibfield  {journal} {\bibinfo  {journal} {J. Phys. Soc. Japan}\ }\textbf
  {\bibinfo {volume} {68}},\ \bibinfo {pages} {683} (\bibinfo {year}
  {1999})}\BibitemShut {NoStop}%
\bibitem [{\citenamefont {Goupalov}\ \emph {et~al.}(2003)\citenamefont
  {Goupalov}, \citenamefont {Lavallard}, \citenamefont {Lamouche},\ and\
  \citenamefont {Citrin}}]{goupalov2003a}%
  \BibitemOpen
  \bibfield  {author} {\bibinfo {author} {\bibfnamefont {S.~V.}\ \bibnamefont
  {Goupalov}}, \bibinfo {author} {\bibfnamefont {P.}~\bibnamefont {Lavallard}},
  \bibinfo {author} {\bibfnamefont {G.}~\bibnamefont {Lamouche}},\ and\
  \bibinfo {author} {\bibfnamefont {D.~S.}\ \bibnamefont {Citrin}},\ }\bibfield
   {title} {\bibinfo {title} {Electrodynamical treatment of the electron-hole
  long-range exchange interaction in semiconductor nanocrystals},\ }\href
  {https://doi.org/10.1134/1.1569020} {\bibfield  {journal} {\bibinfo
  {journal} {Fiz. Tverd. Tela}\ }\textbf {\bibinfo {volume} {45}},\ \bibinfo
  {pages} {730} (\bibinfo {year} {2003})},\ \bibinfo {note} {[Phys. Solid State
  {\bf 2003}, {\it 45}, 768--781]}\BibitemShut {NoStop}%
\bibitem [{\citenamefont {Goupalov}(2003)}]{goupalov2003}%
  \BibitemOpen
  \bibfield  {author} {\bibinfo {author} {\bibfnamefont {S.~V.}\ \bibnamefont
  {Goupalov}},\ }\bibfield  {title} {\bibinfo {title} {Light scattering on
  exciton resonance in a semiconductor quantum dot: Exact solution},\ }\href
  {https://doi.org/DOI: 10.1103/PhysRevB.68.125311} {\bibfield  {journal}
  {\bibinfo  {journal} {Phys. Rev. B}\ }\textbf {\bibinfo {volume} {68}},\
  \bibinfo {pages} {125311} (\bibinfo {year} {2003})}\BibitemShut {NoStop}%
\bibitem [{\citenamefont {Varshalovich}\ \emph {et~al.}(1988)\citenamefont
  {Varshalovich}, \citenamefont {Moskalev},\ and\ \citenamefont
  {Khersonskii}}]{Varshalovich}%
  \BibitemOpen
  \bibfield  {author} {\bibinfo {author} {\bibfnamefont {D.~A.}\ \bibnamefont
  {Varshalovich}}, \bibinfo {author} {\bibfnamefont {A.~N.}\ \bibnamefont
  {Moskalev}},\ and\ \bibinfo {author} {\bibfnamefont {V.~K.}\ \bibnamefont
  {Khersonskii}},\ }\href@noop {} {\emph {\bibinfo {title} {Quantum theory of
  angular momentum}}}\ (\bibinfo  {publisher} {World Scientific},\ \bibinfo
  {year} {1988})\BibitemShut {NoStop}%
\bibitem [{\citenamefont {Gross}\ and\ \citenamefont
  {Haroche}(1982)}]{Gross1982}%
  \BibitemOpen
  \bibfield  {author} {\bibinfo {author} {\bibfnamefont {M.}~\bibnamefont
  {Gross}}\ and\ \bibinfo {author} {\bibfnamefont {S.}~\bibnamefont
  {Haroche}},\ }\bibfield  {title} {\bibinfo {title} {Superradiance: An essay
  on the theory of collective spontaneous emission},\ }\href
  {https://doi.org/10.1016/0370-1573(82)90102-8} {\bibfield  {journal}
  {\bibinfo  {journal} {Phys. Rep.}\ }\textbf {\bibinfo {volume} {93}},\
  \bibinfo {pages} {301} (\bibinfo {year} {1982})}\BibitemShut {NoStop}%
\bibitem [{\citenamefont {Warner}\ \emph {et~al.}(2004)\citenamefont {Warner},
  \citenamefont {Thomsen}, \citenamefont {Watt}, \citenamefont {Heckenberg},\
  and\ \citenamefont {Rubinsztein-Dunlop}}]{Warner2004}%
  \BibitemOpen
  \bibfield  {author} {\bibinfo {author} {\bibfnamefont {J.~H.}\ \bibnamefont
  {Warner}}, \bibinfo {author} {\bibfnamefont {E.}~\bibnamefont {Thomsen}},
  \bibinfo {author} {\bibfnamefont {A.~R.}\ \bibnamefont {Watt}}, \bibinfo
  {author} {\bibfnamefont {N.~R.}\ \bibnamefont {Heckenberg}},\ and\ \bibinfo
  {author} {\bibfnamefont {H.}~\bibnamefont {Rubinsztein-Dunlop}},\ }\bibfield
  {title} {\bibinfo {title} {Time-resolved photoluminescence spectroscopy of
  ligand-capped {PbS} nanocrystals},\ }\href
  {https://doi.org/10.1088/0957-4484/16/2/001} {\bibfield  {journal} {\bibinfo
  {journal} {Nanotechnology}\ }\textbf {\bibinfo {volume} {16}},\ \bibinfo
  {pages} {175} (\bibinfo {year} {2004})}\BibitemShut {NoStop}%
\bibitem [{\citenamefont {Oron}\ \emph {et~al.}(2009)\citenamefont {Oron},
  \citenamefont {Aharoni}, \citenamefont {de~Mello~Donega}, \citenamefont {van
  Rijssel}, \citenamefont {Meijerink},\ and\ \citenamefont {Banin}}]{Oron2009}%
  \BibitemOpen
  \bibfield  {author} {\bibinfo {author} {\bibfnamefont {D.}~\bibnamefont
  {Oron}}, \bibinfo {author} {\bibfnamefont {A.}~\bibnamefont {Aharoni}},
  \bibinfo {author} {\bibfnamefont {C.}~\bibnamefont {de~Mello~Donega}},
  \bibinfo {author} {\bibfnamefont {J.}~\bibnamefont {van Rijssel}}, \bibinfo
  {author} {\bibfnamefont {A.}~\bibnamefont {Meijerink}},\ and\ \bibinfo
  {author} {\bibfnamefont {U.}~\bibnamefont {Banin}},\ }\bibfield  {title}
  {\bibinfo {title} {Universal role of discrete acoustic phonons in the
  low-temperature optical emission of colloidal quantum dots},\ }\href
  {https://doi.org/10.1103/physrevlett.102.177402} {\bibfield  {journal}
  {\bibinfo  {journal} {Physical Review Letters}\ }\textbf {\bibinfo {volume}
  {102}},\ \bibinfo {pages} {117402} (\bibinfo {year} {2009})}\BibitemShut
  {NoStop}%
\bibitem [{\citenamefont {Kigel}\ \emph {et~al.}(2009)\citenamefont {Kigel},
  \citenamefont {Brumer}, \citenamefont {Maikov}, \citenamefont {Sashchiuk},\
  and\ \citenamefont {Lifshitz}}]{Kigel2009}%
  \BibitemOpen
  \bibfield  {author} {\bibinfo {author} {\bibfnamefont {A.}~\bibnamefont
  {Kigel}}, \bibinfo {author} {\bibfnamefont {M.}~\bibnamefont {Brumer}},
  \bibinfo {author} {\bibfnamefont {G.}~\bibnamefont {Maikov}}, \bibinfo
  {author} {\bibfnamefont {A.}~\bibnamefont {Sashchiuk}},\ and\ \bibinfo
  {author} {\bibfnamefont {E.}~\bibnamefont {Lifshitz}},\ }\bibfield  {title}
  {\bibinfo {title} {The ground-state exciton lifetime of {PbSe} nanocrystal
  quantum dots},\ }\href {https://doi.org/10.1016/j.spmi.2008.11.026}
  {\bibfield  {journal} {\bibinfo  {journal} {Superlattices and
  Microstructures}\ }\textbf {\bibinfo {volume} {46}},\ \bibinfo {pages} {272}
  (\bibinfo {year} {2009})}\BibitemShut {NoStop}%
\bibitem [{\citenamefont {Biadala}\ \emph {et~al.}(2009)\citenamefont
  {Biadala}, \citenamefont {Louyer}, \citenamefont {Tamarat},\ and\
  \citenamefont {Lounis}}]{Biadala2009}%
  \BibitemOpen
  \bibfield  {author} {\bibinfo {author} {\bibfnamefont {L.}~\bibnamefont
  {Biadala}}, \bibinfo {author} {\bibfnamefont {Y.}~\bibnamefont {Louyer}},
  \bibinfo {author} {\bibfnamefont {P.}~\bibnamefont {Tamarat}},\ and\ \bibinfo
  {author} {\bibfnamefont {B.}~\bibnamefont {Lounis}},\ }\bibfield  {title}
  {\bibinfo {title} {Direct observation of the two lowest exciton zero-phonon
  lines in single cdse/zns nanocrystals},\ }\bibfield  {journal} {\bibinfo
  {journal} {Physical Review Letters}\ }\textbf {\bibinfo {volume} {103}},\
  \href {https://doi.org/10.1103/physrevlett.103.037404}
  {10.1103/physrevlett.103.037404} (\bibinfo {year} {2009})\BibitemShut
  {NoStop}%
\bibitem [{\citenamefont {Poddubny}\ \emph {et~al.}(2012)\citenamefont
  {Poddubny}, \citenamefont {Nestoklon},\ and\ \citenamefont
  {Goupalov}}]{Poddubny12}%
  \BibitemOpen
  \bibfield  {author} {\bibinfo {author} {\bibfnamefont {A.~N.}\ \bibnamefont
  {Poddubny}}, \bibinfo {author} {\bibfnamefont {M.~O.}\ \bibnamefont
  {Nestoklon}},\ and\ \bibinfo {author} {\bibfnamefont {S.~V.}\ \bibnamefont
  {Goupalov}},\ }\bibfield  {title} {\bibinfo {title} {Anomalous suppression of
  valley splittings in lead salt nanocrystals without inversion center},\
  }\href {https://doi.org/10.1103/PhysRevB.86.035324} {\bibfield  {journal}
  {\bibinfo  {journal} {Phys. Rev. B}\ }\textbf {\bibinfo {volume} {86}},\
  \bibinfo {pages} {035324} (\bibinfo {year} {2012})}\BibitemShut {NoStop}%
\bibitem [{\citenamefont {Koster}\ \emph {et~al.}(1963)\citenamefont {Koster},
  \citenamefont {Dimmock}, \citenamefont {Wheeler},\ and\ \citenamefont
  {Statz}}]{Koster}%
  \BibitemOpen
  \bibfield  {author} {\bibinfo {author} {\bibfnamefont {G.~F.}\ \bibnamefont
  {Koster}}, \bibinfo {author} {\bibfnamefont {J.~O.}\ \bibnamefont {Dimmock}},
  \bibinfo {author} {\bibfnamefont {R.~G.}\ \bibnamefont {Wheeler}},\ and\
  \bibinfo {author} {\bibfnamefont {H.}~\bibnamefont {Statz}},\ }\href@noop {}
  {\emph {\bibinfo {title} {The Properties of the Thirty-Two Point Groups}}}\
  (\bibinfo  {publisher} {M.I.T. Press, Cambridge},\ \bibinfo {year}
  {1963})\BibitemShut {NoStop}%
\bibitem [{\citenamefont {Allan}\ and\ \citenamefont
  {Delerue}(2004)}]{Delerue2004b}%
  \BibitemOpen
  \bibfield  {author} {\bibinfo {author} {\bibfnamefont {G.}~\bibnamefont
  {Allan}}\ and\ \bibinfo {author} {\bibfnamefont {C.}~\bibnamefont
  {Delerue}},\ }\bibfield  {title} {\bibinfo {title} {{C}onfinement effects in
  {P}b{S}e quantum wells and nanocrystals},\ }\href
  {https://doi.org/10.1103/PhysRevB.70.245321} {\bibfield  {journal} {\bibinfo
  {journal} {Phys. Rev. B}\ }\textbf {\bibinfo {volume} {70}},\ \bibinfo
  {pages} {245321} (\bibinfo {year} {2004})}\BibitemShut {NoStop}%
\bibitem [{\citenamefont {Hu}\ \emph {et~al.}(2019)\citenamefont {Hu},
  \citenamefont {Kim}, \citenamefont {Krishnamurthy}, \citenamefont {Avdeev},
  \citenamefont {Nestoklon}, \citenamefont {Singh}, \citenamefont {Malko},
  \citenamefont {Goupalov}, \citenamefont {Hollingsworth},\ and\ \citenamefont
  {Htoon}}]{Hu19}%
  \BibitemOpen
  \bibfield  {author} {\bibinfo {author} {\bibfnamefont {Z.}~\bibnamefont
  {Hu}}, \bibinfo {author} {\bibfnamefont {Y.}~\bibnamefont {Kim}}, \bibinfo
  {author} {\bibfnamefont {S.}~\bibnamefont {Krishnamurthy}}, \bibinfo {author}
  {\bibfnamefont {I.~D.}\ \bibnamefont {Avdeev}}, \bibinfo {author}
  {\bibfnamefont {M.~O.}\ \bibnamefont {Nestoklon}}, \bibinfo {author}
  {\bibfnamefont {A.}~\bibnamefont {Singh}}, \bibinfo {author} {\bibfnamefont
  {A.~V.}\ \bibnamefont {Malko}}, \bibinfo {author} {\bibfnamefont {S.~V.}\
  \bibnamefont {Goupalov}}, \bibinfo {author} {\bibfnamefont {J.~A.}\
  \bibnamefont {Hollingsworth}},\ and\ \bibinfo {author} {\bibfnamefont
  {H.}~\bibnamefont {Htoon}},\ }\bibfield  {title} {\bibinfo {title} {Intrinsic
  exciton photophysics of {PbS} quantum dots revealed by low-temperature single
  nanocrystal spectroscopy},\ }\href
  {https://doi.org/10.1021/acs.nanolett.9b02937} {\bibfield  {journal}
  {\bibinfo  {journal} {Nano Lett.}\ }\textbf {\bibinfo {volume} {19}},\
  \bibinfo {pages} {8519} (\bibinfo {year} {2019})}\BibitemShut {NoStop}%
\end{thebibliography}%

\end{document}